\documentclass[preprint,12pt]{elsarticle}

\usepackage{amssymb}
\usepackage{amsmath}
\usepackage{amsthm}

\usepackage{pdflscape}
\usepackage{geometry}
\usepackage{caption}

\usepackage{booktabs,threeparttable}

\usepackage{algorithmicx}

\usepackage{multirow}

\journal{Combustion and Flame}

\begin{document}

\begin{frontmatter}



\title{ A species-clustered ODE solver for large-scale chemical kinetics using detailed mechanisms } 

 \author[label1]{Jian-Hang Wang}
 \author[label1]{Shucheng Pan} 
 \author[label1]{Xiangyu Y. Hu\corref{xiangyu}}
 \author[label1]{Nikolaus A. Adams}
 \cortext[xiangyu]{Corresponding author}
 \address[label1]{Chair of Aerodynamics and Fluid Mechanics, Department of Mechanical Engineering, Technical University of Munich, 85748 Garching, Germany}

\begin{abstract} 

In this study, a species-clustered ordinary differential equations (ODE) solver for chemical kinetics with large detailed mechanisms based on operator-splitting is presented. The ODE system is split into clusters of species by using graph partition methods which has been intensively studied in areas of model reduction, parameterization and coarse-graining, etc. , such as diffusion maps based on the concept of Markov random walk. Definition of the weight (similarity) matrix is application-driven and according to chemical kinetics. 
Each cluster of species is then integrated by VODE, an implicit solver which is intractable and costly for large systems of many species and reactions. Expected speedup in computational efficiency is observed by numerical experiments on three zero-dimensional (0D) auto-ignition problems, considering the detailed hydrocarbon/air combustion mechanisms in varying scales, from 53 species with 325 reactions of methane to 2115 species with 8157 reactions of n-hexadecane.
             
\end{abstract}


\end{frontmatter}




\section{Introduction}

Gasline, diesel and jet fuels particularly those derived from petroleum sources are composed of hundreds of compounds \cite{MERSIN2014132}. As the hydrocarbon species grows, so does the size of kinetic mechanism to model hydrocarbon oxidation. For example, the detailed mechanism for methyl decanoate, a biomass fuel surrogate, consists of 3036 species and 8555 reactions \cite{herbinet2008detailed,LU2009192}. From the aspect of accurate prediction of combustion processes such as ignition, extinction and flame propagation, large-scale chemical kinetics is without a question important \cite{xu2016sparse}. However, readily adoption of a comprehensive detailed reaction mechanism for computational simulation is limited by the current computing power. Computation of the aforementioned mechanism is time consuming even for 0D simulations \cite{LU2009192}, no matter using explicit or implicit solvers. This limitation therefore contributes to the development of mechanism reduction methods, e.g. directed relation graph (DRG) based methods \cite{lu2005directed,PEPIOTDESJARDINS200867,SUN20101298,NIEMEYER20101760}, etc.      

Moreover, severe chemical stiffness generally exists due to dramatic differences in the varying species and reaction timescales, so that the high-cost implicit ODE solvers, e.g. VODE \cite{brown1989vode} and DASAC \cite{CARACOTSIOS1985359}, which allow the robust use of reasonably large timesteps, are typically required for time integration of combustion systems \cite{xu2016sparse}. Since Jacobian evaluation and factorization in implicit solvers dominate the computational cost, the scaling of CPU time over the number of species in the mechanism is roughly from $O(N^2)$ to $O(N^3)$ with dense matrix operations \cite{perini2014study,damian2002kinetic}. This disadvantage severely hinders the computation of large-size mechanisms.       

For general multi-dimensional reactive flows, operator splitting has been widely used to separate chemistry integration from that of transport processes to reduce computational efforts \cite{ren2014dynamic,singer2004exploiting,singer2006operator,knio1999semi,ren2008second}. C. Xu \cite{xu2016sparse} and Y. Gao \cite{gao2015dynamic} adaptively separate the dynamic system into a fast operator including only fast reactions and a slow operator including slow reactions and the transport process, with each part being imposed of an implicit solver and a more efficient explicit solver, respectively. For the chemical dynamics only, K. Nguyen \cite{nguyen2009mass} aiming at preserving mass conservation and definite positivity solves exactly each chemical reaction after splitting the multi-reaction system into decoupled processes. Pan \textit{et al.} \cite{pan2018network} introduce the graph/network partition into large-scale stochastic and mass concentration based chemical networks.           

The quadric/cubic scaling of CPU time to mechanism size using implicit ODE solvers implies the computational cost of solving a sequence of smaller subsystems ought to be much less than that of solving the entire system in one operation. Therefore, unlike the above use of operator splitting in decoupling two or more physical processes, we start with splitting the large-scale chemical kinetics in terms of the involved species. Once the participating species in the large mechanism have been clustered into subsets of a smaller and equal size, an implicit solver can be applied to each group with the reduced matrix scale. To minimize the splitting error, diffusion maps \cite{coifman2006diffusion,lafon2006diffusion,coifman2005geometric} are utilized to analyze the pairwise interaction relations of species by constructing a weight or similarity matrix in the scenario of chemical kinetics, such that intensively interacting and mutually dependent species can be clustered into the same group. To partition the species into equal clusters, a balanced k-means algorithm \cite{malinen2014balanced} is needed in association.          

The paper is organized as follows. In Section \ref{Methodology}, we introduce the ODE system of chemical kinetics and formulate the species-clustered solver illustrated by a simple model example. Results from the proposed method for three detailed mechanisms in varying scales are presented and discussed in Section \ref{Results}, considering the 0D auto-ignition problem at constant-volume and adiabatic conditions. Conclusions are drawn in Section \ref{Conclusions}.

\section{Methodology}
\label{Methodology}

\subsection{Operator splitting by species for chemical kinetics}

The ODE system of chemical kinetics under adiabatic and constant-volume conditions can be expressed as 
\begin{equation}\label{S_r}
\begin{aligned}
\frac{dy_i}{dt} = \frac{\dot{\omega_i}}{\rho},\quad i=1,\dots,N_s,
\end{aligned}
\end{equation} 
where $y_i$ and $w_i$ denote the mass fraction and the total production rate of species $i$, respectively, in the mechanism consisting of $N_s$ species and $N_r$ reactions. Each reaction can be generally written as
\begin{equation}\label{reaction}
\sum_{i=1}^{N_s} \nu_{ji}^f X_i  \Longleftrightarrow \sum_{i=1}^{N_s} \nu_{ji}^b X_i,  \quad j=1,\dots,N_r,
\end{equation} 
where $\nu_{ji}^f$ and $\nu_{ji}^b$ are the stoichiometric coefficients of species $i$ appearing as a reactant and as a product in reaction $j$. The total production rate of species $i$ in Eq. \eqref{S_r} is the sum of the production rate from each single elementary reaction by
\begin{equation}
\dot{\omega_i} = W_i \sum_{j=1}^{N_r} (\nu_{ji}^b-\nu_{ji}^f) \left[ k_j^f \prod_{l=1}^{N_s} \left[\frac{\rho_l}{W_l}\right]^{\nu_{jl}^f} - k_j^b \prod_{l=1}^{N_s} \left[\frac{\rho_l}{W_l}\right]^{\nu_{jl}^b}  \right],
\end{equation} 
with $k_j^f$ and $k_j^b$ denoting the forward and backward reaction rates of each chemical reaction and $W_i$ being the molecular weight of the $i^{th}$ species.  
With fixed total density and constant specific internal energy, the equation of state (EoS) for ideal gas mixture can be used to determine the evolution of mixture temperature and thus to close the system.
 
Assuming that we have a variable vector $\Phi = \{y_1,\cdots,y_{N_s}\}^T$ at time level $n$, to integrate the above ODE system for one timestep of $\Delta t$, the implicit solver VODE \cite{brown1989vode} is employed in the following form,
\begin{equation}\label{VODE}
\begin{aligned}
\Phi^{n+1} = R_{\Delta t}( \Phi^n ),
\end{aligned}
\end{equation}
with operator $R$ representing the time integration of VODE.    
Using operator splitting by species upon Eq. \eqref{VODE}, we can obtain 
\begin{equation}\label{VODE1}
\begin{aligned}
\Phi^{n+1} = R_{\Delta t}( \Phi_1^n ) \circ R_{\Delta t}( \Phi_2^n ) \cdots R_{\Delta t}( \Phi_N^n ),
\end{aligned}
\end{equation}   
corresponding the Lie-Trotter splitting scheme \cite{mclachlan2002splitting}, where $\Phi_{k, k=1,2,\dots,N}$ denotes the mass fractions of the species clustered in subset $S_k$ out of $N$ subsets in total. Clustering of species in each subset should be subject to 
\begin{equation}\label{partitioning}
\begin{aligned}
\Phi & = \{ \Phi_1, \cdots, \Phi_N \}^T, \\
S & = \cup_{k=1}^{N} S_k, \ S_i \cap S_j = \emptyset \ \text{if} \ i \neq j.
\end{aligned}
\end{equation} 
Note that each cluster/subset of species should have no overlapping parts with others and an equal number of species in each subset is presumed with at most one species more or less (which requests a balanced partition/clustering algorithm \cite{malinen2014balanced}). 
The extension to higher-order splitting of Strang \cite{strang1968construction} is straightforward but inevitably more time-consuming. Recalling that the scaling of computational cost to the number of species or the size of the kinetic mechanism involved using an implicit solver such as VODE is \cite{xu2016sparse},
\begin{equation}\label{time_cost}
\begin{aligned}
t_{\text{CPU}} \sim  O(N_s^2) \ \text{to} \ O(N_s^3),
\end{aligned}
\end{equation}           
the total cost after splitting by species can be roughly reduced to
\begin{equation}\label{time_cost1}
\begin{aligned}
t^{'}_{\text{CPU}} \sim  \frac{1}{N} O(N_s^2) \ \text{to} \  \frac{1}{N^2} O(N_s^3).
\end{aligned}
\end{equation} 
As a result, take an extremely large mechanism consisting of ten thousand species for example, if we split the system into ten clusters with the Lie-Trotter scheme, the computational speedup will be ten to a hundred times ideally, which is attractive without appealing to additional sparse matrix techniques \cite{perini2014study,schwer2002upgrading,damian2002kinetic}. 

The essence of operator splitting by species for chemical kinetics lies in clustering species into subsets, each corresponding to a sub-ODE-system to be integrated by VODE or other implicit solvers. The merits of operator splitting by species include the speedup of computational efficiency regarding the same implicit solver as well as being quickly convergent and very numerically stable \cite{pan2018network}. Also, the speedup factor mainly ascribes to the number of equal-sized clusters by splitting. i.e. $N$ in Eq. \eqref{partitioning}.                 

\subsection{Graph-based species clustering}

A chemical reaction system with multiple species and reactions can be actually translated to be a bi-partite graph \cite{domijan2008some}, in which two sets of nodes representing the chemical species and reactions, respectively, are involved. Herein, we simply consider a finite graph consisting of the chemical species only and the non-linear coupling between pairs of species through reactions is abstracted into undirected edges linking every two nodes of species. For the sake of illustration, without loss of generality, we start with a model kinetics consisting of six species, $\{A,B,C,D,E,F\}$, and six first-order one-way reactions, i.e.
\begin{equation}\label{model_kinetcis}
\begin{aligned}
A \xrightarrow{k_1} C,\
B \xrightarrow{k_2} C,\
C \xrightarrow{k_3} B,\\
D \xrightarrow{k_4} C,\
E \xrightarrow{k_5} D,\
F \xrightarrow{k_6} D,\\
\end{aligned}
\end{equation} 
where $k$ is the constant reaction rate. Analytical exact solution for the model kinetics can be easily obtained using symbolic computations of MATLAB\textsuperscript{\textregistered} \cite{MATLAB}. 
   
\begin{figure}
  \centering
  \includegraphics[trim={2cm 1cm 0cm 3cm},scale=0.5]{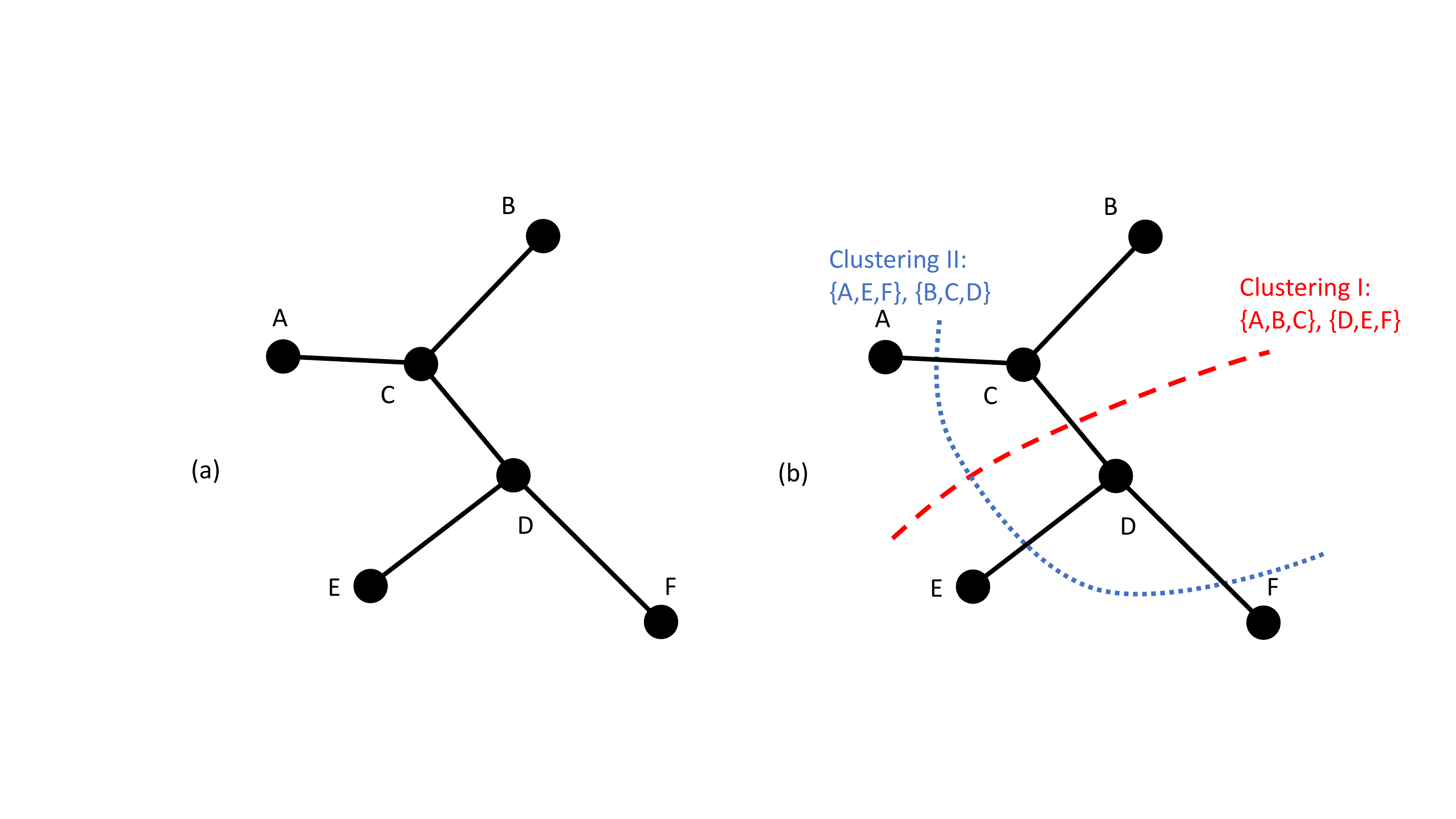} \\
  \includegraphics[trim={2cm 1cm 0cm 3cm},scale=0.5]{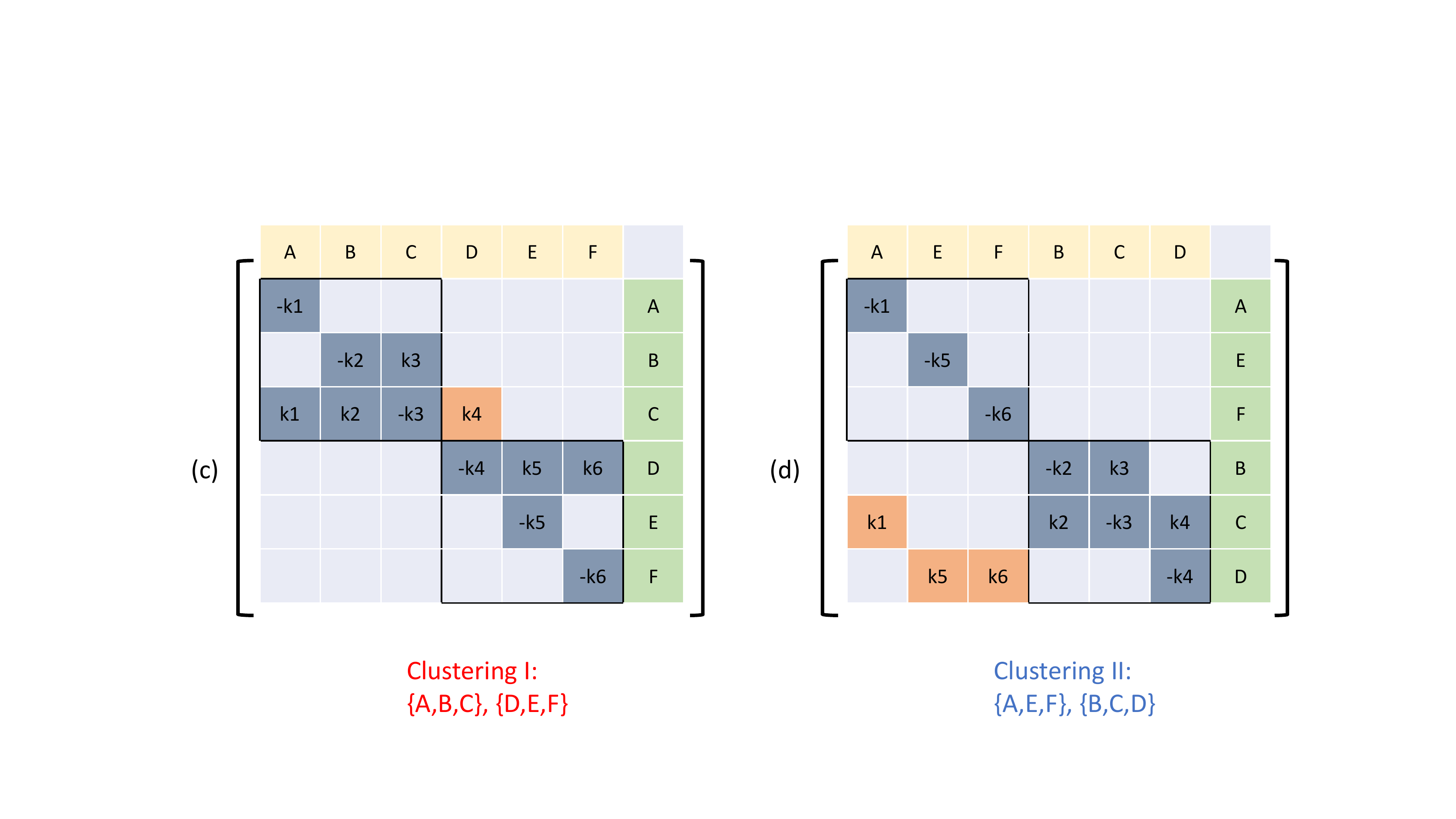} \\
  \caption{ Model kinetics example for species clustering. (a) Each node represents one species in $\{A,B,C,D,E,F\}$ and edges, say $e(A,C)$, indicates the linked two species jointly participate in at least one reaction as reactant or product; (b) Two equal-sized clusterings are easily obtained as $(\{A,B,C\},\{D,E,F\})$ and $(\{A,E,F\},\{B,C,D\})$ by cutting off corresponding edges; (c) Rearranged Jacobian matrix in the order of clustering I; (d) Rearranged Jacobian matrix in the order of clustering II. }    
  \label{schematic1}
\end{figure}

\begin{figure}
	\centering
	\includegraphics[scale=0.32]{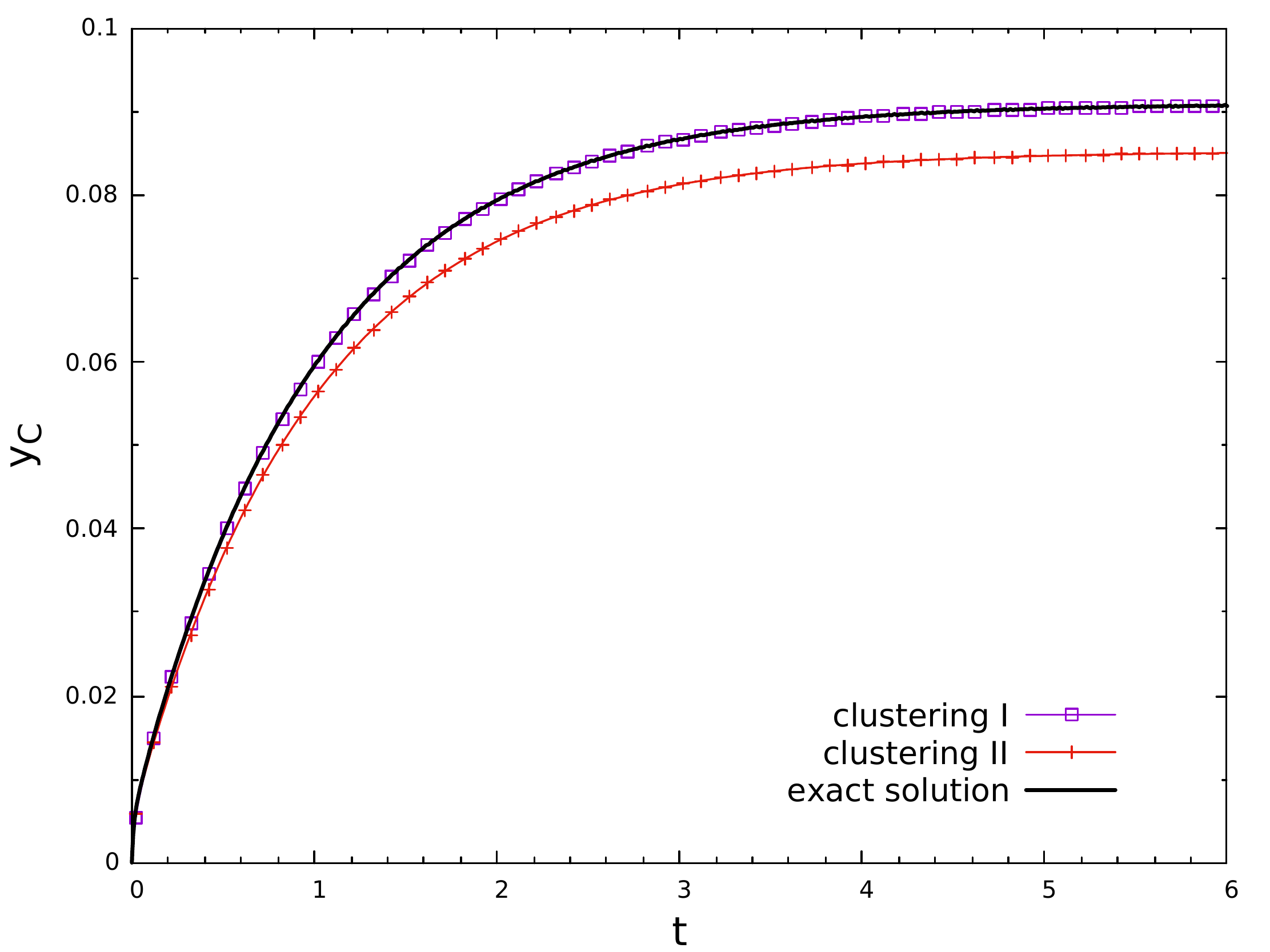}
	\includegraphics[scale=0.32]{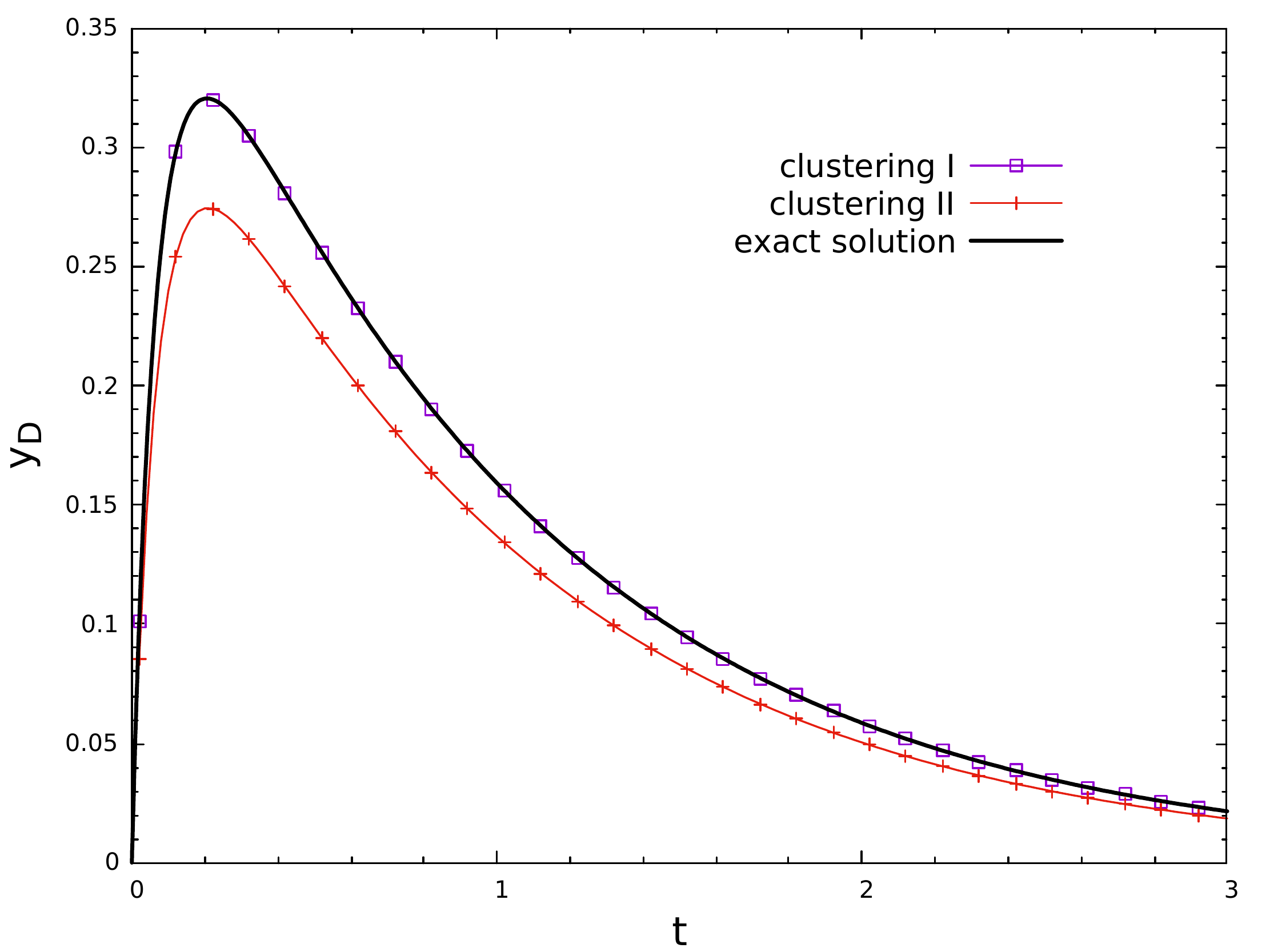} 
	\caption{ Numerical integration results with two clusterings by Lie-Trotter splitting scheme and compared with the analytical exact solution. Reaction rates are $k_1=1,k_2=10,k_3=100,k_4=1,k_5=10,k_6=20$, and the initial condition is $y_A=0.6,y_E=0.2,y_F=0.2$ with zero mass fractions of $B,C,D$. Computational timestep is fixed at $\Delta t=0.02$. }    
	\label{schematic2}
\end{figure}

We firstly construct the graph of species as shown in Fig. \ref{schematic1}(a). Based on the graph, we can easily obtain two clusterings I and II with two subsets (i.e. $N=2$). Regarding Clustering I in \ref{schematic1}(b), we cut off the link between species $C$ and $D$ by observing the dumbbell-like structure in the graph such that the strong couplings inside $\{A,B,C\}$ and $\{D,E,F\}$ both can be preserved. In contrast, we cluster the loosely coupled $\{A,E,F\}$ together and leave the rest to compose the other cluster, as in Clustering II. In particular, the distance in the graph between $\left(A,E\right)$ or $\left(A,F\right)$ is remote, separated by at other two species. The difference of two clusterings can be also reflected in the rearranged Jacobian matrices in the order of splitting and clustering as shown in Fig. \ref{schematic1}(c) and (d). We can see that in the first clustering situation, when solving the cluster of ${A,B,C}$ first, only the impact of species $C$ is considered as constant since $k_4$ is outside the sub-Jacobian matrix. And when solving the other cluster of $\{D,E,F\}$ subsequently, impacts of species $A$, $B$ and $C$ is inactive due to the corresponding zero entries, such that the accuracy of solving $\{D,E,F\}$ here is identical as in non-split matrix operations. In total, the splitting error mainly comes from only one location/element in the matrix, i.e. the $k_4$ block in brick-red color in Fig. \ref{schematic1}(c). In terms of the scenario of Clustering II, although solving the first cluster of $\{A,E,F\}$ introduces no error owing to the zero entries behind the sub-matrix, non-trivial errors will be brought in when solving the following cluster, $\{B,C,D\}$, by simply considering $k_1 y_A$ for the production of species $C$ and $k_5 y_E + k_6 y_F$ for the production of species $D$ to be constant. We carry out a numerical test by solving the model kinetics in operator splitting manner using the two clusterings, in Fig. \ref{schematic2}. Analytical exact solution is also presented for comparison. We can readily see that the solution based on Clustering I agrees quite well with the exact solution, while the Clustering II solution underestimates both the mass fractions of species $C$ and $D$. This observation is also agreeable with the previous discussion about the split operations, which are undertaken upon each sub-matrix dynamically but the outside elements statically. Therefore, the quality of clustering is supposed to own an important influence on the splitting error and thus the accuracy of our integration.                          

Given the prescribed number of clusters, there are tens or hundreds clusterings according to the combination theory. One simple case-independent way to cluster the species is according to the indexes of species appearing in the mechanism, without either personal experience or beforehand knowledge of the structure of the given mechanism. 
In order to improve the quality of clustering which is highly related to the splitting error, one promising idea is to cluster the 'close' nodes with each other into the same subset, which indicates that for chemical kinetics it is favorable to have the species with strong couplings or interactions in the same cluster. In this paper, we introduce diffusion maps \cite{coifman2005geometric,coifman2006diffusion,lafon2006diffusion}, a non-linear technique for dimensionality reduction, data set parameterization and clustering, to serve the purpose.        
 
Let $G=(\Omega,W)$ be a finite graph of $n$ nodes, where the weight matrix $W=\{ w(x,y) \}_{x,y\in \Omega}$ should satisfy conditions of symmetry and pointwise positivity \cite{lafon2006diffusion}. As it is application-driven, the definition of weight matrix needs to reflect the degree of similarity or affinity of nodes $x$ and $y$. Diffusion maps start with the user-defined weight matrix which may be considered as a measure of the local geometry and utilize the idea of Markov random walk to describe the connectivity of nodes to be a diffusion process. As the diffusion progresses, it integrates local geometry to reveal geometric structures of the data set. We herein simply skip the technical details of diffusion maps and focus our concentration on species clustering using diffusion maps for chemical kinetics. 

\begin{figure}
	\centering
	\includegraphics[scale=0.45]{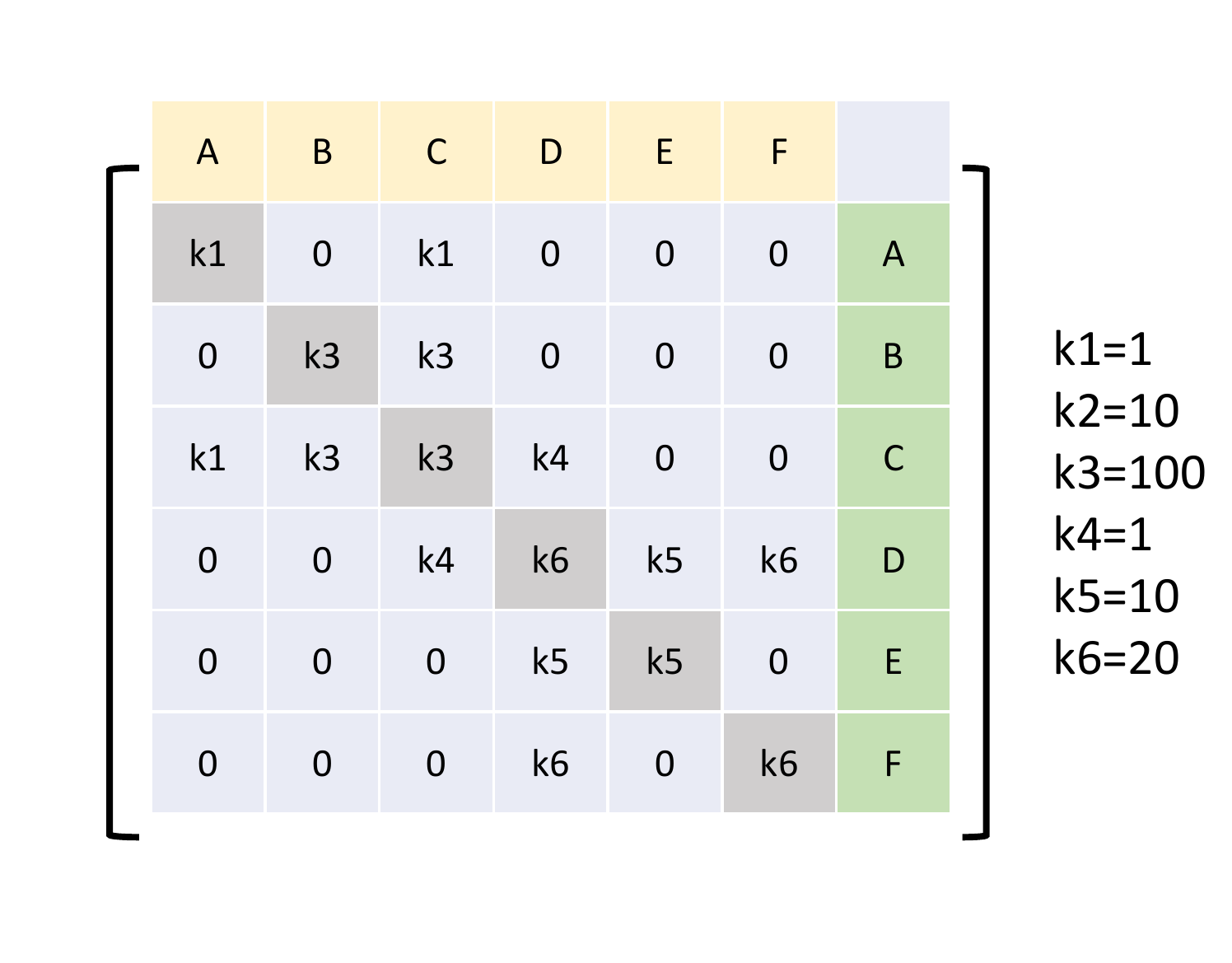}
	\includegraphics[trim={2cm 1cm 3cm 0cm},scale=0.3]{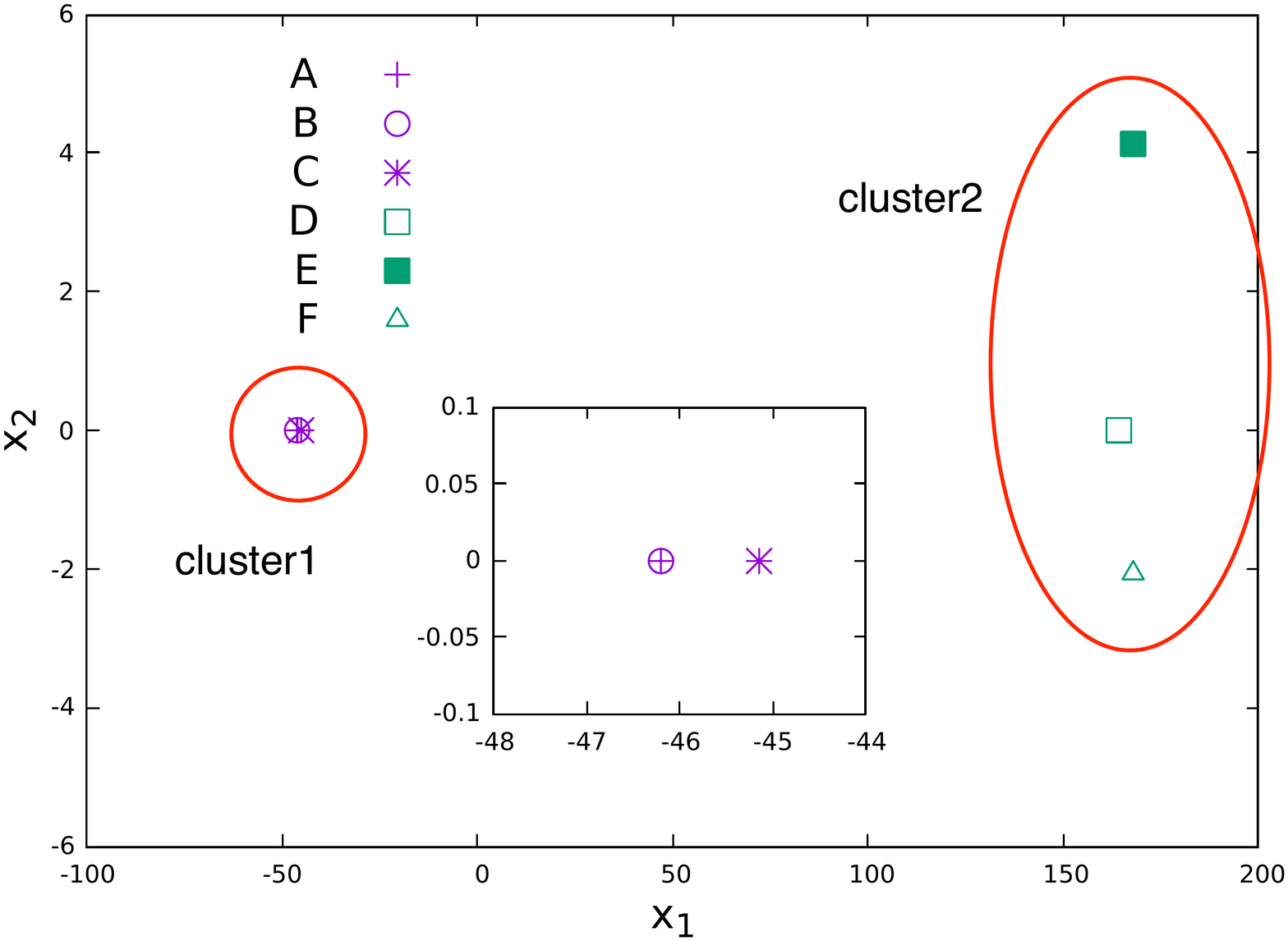}
	\caption{ Weight matrix of diffusion maps for the model kinetics (left); embedding and clustering of species in 2D diffusion space (right) }    
	\label{weight_matrix_fig}
\end{figure}

For the model kinetics, with the help of species graph in Fig. \ref{schematic1}(a), we define the weight matrix $W$ by 
\begin{equation}\label{weight_matrix}
\begin{aligned}
 w(x,y) =
\begin{cases}
 max(k_j), & \text{if} \ x \ \text{and} \ y \ \text{both participate in reaction} \ j, \\
 \epsilon, & \text{otherwise},
\end{cases} 
\end{aligned}
\end{equation}   
where $\epsilon$ takes a small positive value to avoid zero entries, e.g. $\epsilon=10^{-12}$. 
Especially, the diagonal elements in the weight matrix, $w(x,x)$, can be consistently defined as 
\begin{equation}\label{weight_matrix1}
\begin{aligned}
w(x,x) = max(w(x,y)_{y \neq x }).   
\end{aligned}
\end{equation}
In combination with the reaction rates given in Fig. \ref{schematic2}, the weight matrix obtained by the above definition is shown in Fig. \ref{weight_matrix_fig}. Using diffusion maps to analyze the graph based on our defined weight matrix, we can project the set of species into a diffusion space with at most $n$ dimensions, where the pairwise distance can reveal the connectivity between two species. In Fig. \ref{weight_matrix_fig}, it is shown that species are projected onto a $x_1x_2$ plane using the first two dimensions of the diffusion space. We can see that species $A$, $B$ and $C$ almost collapse into one point and locations of species $D$, $E$ and $F$ in the $x_1$ direction (which is also the first and main dimension) are very close to each other. Their coordinates in the second dimension separate the three species apart. However, the centroids of subset, $\{A,B,C\}$, and subset, $\{D,E,F\}$, are very far from each other. Accordingly, a straightforward clustering using the k-means algorithm (setting $k \equiv N=2$) can be easily obtained, i.e. $(\{A,B,C\},\{D,E,F\})$. This clustering from diffusion maps is the same as the previous Clustering I, indicating that it is an optimal case in two clusters for the model kinetics with reduced splitting errors.               

For much more complicated realistic chemical kinetics especially that involves fuel combustion mechanisms, reaction rates are not always constant but dependent on temperature even pressure of the mixture. This normally can be expressed by the finite-rate Arrhenius model \cite{mott2001chemeq2,wang2018split} and thus the weight matrix as above should also take into account the varying reaction rates with temperature. Rather than sampling at one temperature value, e.g. the initial temperature of an auto-ignition problem of combustible gas mixtures, we take many temperature samples in order to construct a representative weight matrix. Then the derived clustering by diffusion maps based on such a weight matrix can be stored and used for other conditions as long as the same mechanism is involved. In such way, determination of the weight matrix can be treated as a preprocessing step instead of the costly on-the-fly clustering. As multiple scales of the absolute reaction rates exist, usually spanning several orders of magnitude, logarithmic scaling of the reaction rates is performed to avoid underestimating the slow reactions. Also, normalization in each line of the matrix relative to the diagonal species is carried out as
\begin{equation}\label{normalization}
\begin{aligned}
w(x,y) = \frac{w(x,y)}{w(x,x)},   
\end{aligned}
\end{equation}   
and  
\begin{equation}\label{symmetry}
\begin{aligned}
w(x,y) = max(w(x,y),w(y,x))   
\end{aligned}
\end{equation} 
for all species pairs is further executed to guarantee the symmetry of weight matrix in diffusion maps.

\section{Numerical results and discussion}
\label{Results}

In this section of numerical experiments, we consider three detailed mechanisms for hydrocarbon fuel combustion: the GRI-Mech 3.0 mechanism for methane (CH$_4$) \cite{smith1999gri}, the n-heptane (n-C$_7$H$_{16}$) mechanism (Version 2) \cite{curran1998comprehensive,curran2002comprehensive} and the n-hexadecane (n-C$_{16}$H$_{34}$) mechanism \cite{westbrook2007detailed}. The sizes of three mechanisms are listed in Table \ref{mechanism}, with increasing numbers of species and reactions as well as growing computational complexity of time integration. Zero-dimensional auto-ignition of the fuel/air mixture under adiabatic and constant-volume conditions is taken into consideration. 

\begin{table}
\caption{Numbers of species and reactions in detailed mechanisms.}
\centering
\label{mechanism}
{
\begin{tabular}{lcc}
\hline
    			& No. of species	& No. of reactions	\\
\hline	
 CH$_4$			&	53		&	325		\\
 n-C$_7$H$_{16}$ 	&	561		&	2539	\\
 n-C$_{16}$H$_{34}$ &	2115	&	8157	\\
\hline
\end{tabular}
}
\end{table}


\subsection{Methane/air auto-igniton}

The first example takes the ignition delay problem of methane/air mixture. Two initial conditions \cite{seery1970experimental} are considered as in Table \ref{case12}. For Case 1, computation is carried out till $t=0.001$ s and the base timestep is fixed at $\Delta t = 1 \times 10^{-7}$ s (this base timestep is also adopted for other cases without special statements). Computation for Case 2 is till $t=2\times 10^{-4}$ s. CHEMEQ2 \cite{mott2001chemeq2} as a popular explicit ODE solver for chemical kinetics is also employed here for reference both in computational efficiency and numerical accuracy, together with the pure implicit solver VODE without splitting by species. In CHEMEQ2, the convergence parameter of the predictor-corrector method is $1\times10^{-4}$. In VODE, the relative and absolute error thresholds (RTOL and ATOL) are $1 \times 10^{-5}$ and $1 \times 10^{-13}$, respectively.
Since size of the methane mechanism is relatively small, we take into account clustering the 53 species into two subsets and each cluster of species is integrated by VODE in a fractional step manner as in Eq. \eqref{VODE1}. Accuracy and convergence of the splitting method using species clustering are mainly examined in this example. Benefits of computational efficiency from operator splitting by species clustering is to be tested in the following two mechanisms of much larger scales.
As an important parameter to measure the accuracy of both the mechanism and ODE solver, ignition delay times, $t_{ign}$, for the two cases can be referred to \cite{seery1970experimental}, i.e. $t_{ign}=666$ ms for Case 1 and $t_{ign}=110$ ms for Case 2, approximately.

\begin{table}
	\caption{Initial conditions for methane/air mixture.}
	\centering
	\label{case12}
	{
		\begin{tabular}{lccc}
			\hline
					&	CH$_4$-O$_2$-Ar molar ratio	& Temperature (K)	& Pressure (atm) \\
			\hline	
			Case 1	& \multirow{2}{*}{9.1\%-18.2\%-72.7\%}	&	1500		&	1.8		\\
			Case 2 	& 										&	1700		&	2.04	\\
			\hline
		\end{tabular}
	}
\end{table}

\begin{figure}
	\centering
	\includegraphics[scale=0.32]{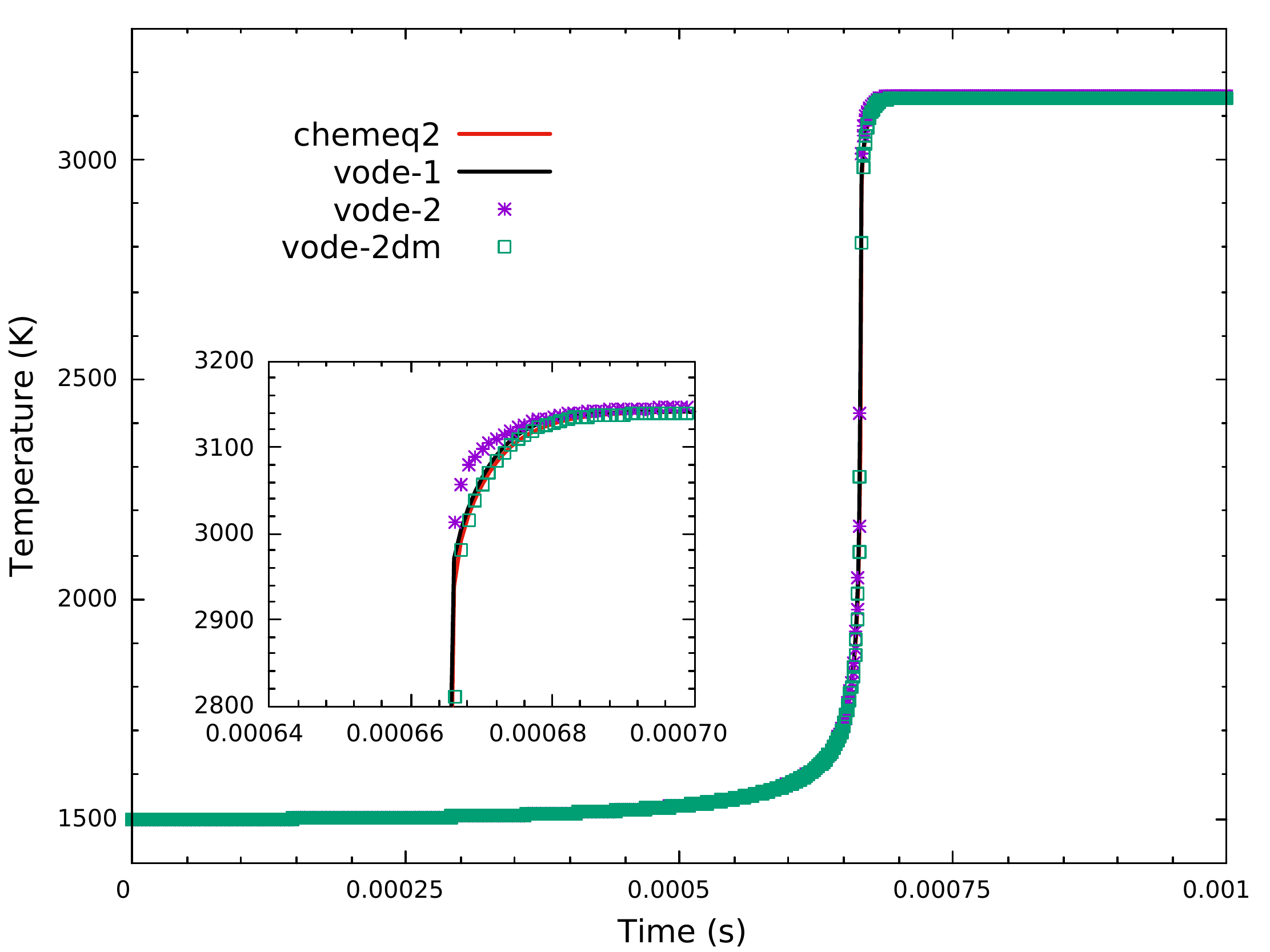}
	\includegraphics[scale=0.32]{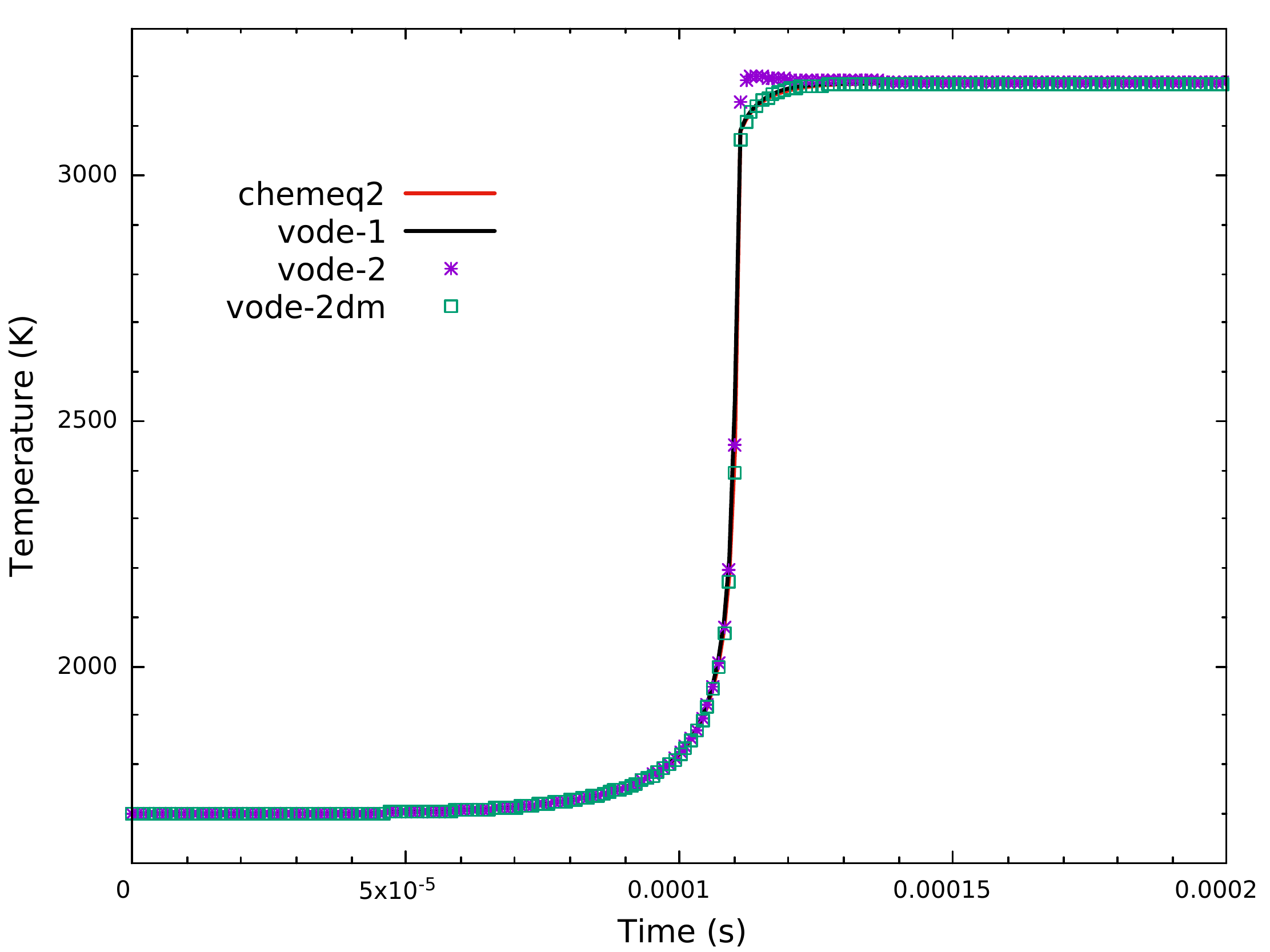} 
	\includegraphics[scale=0.32]{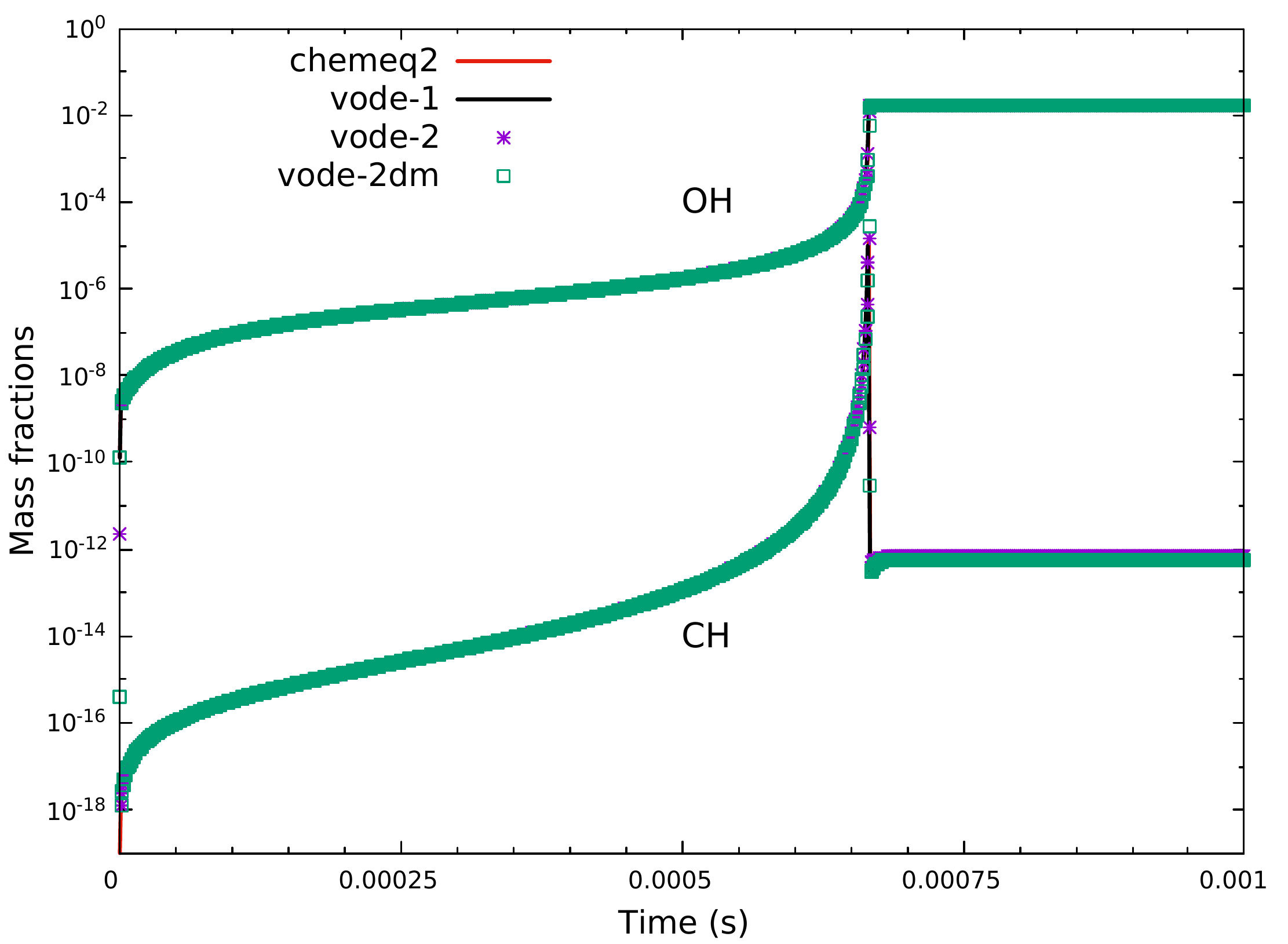}
	\includegraphics[scale=0.32]{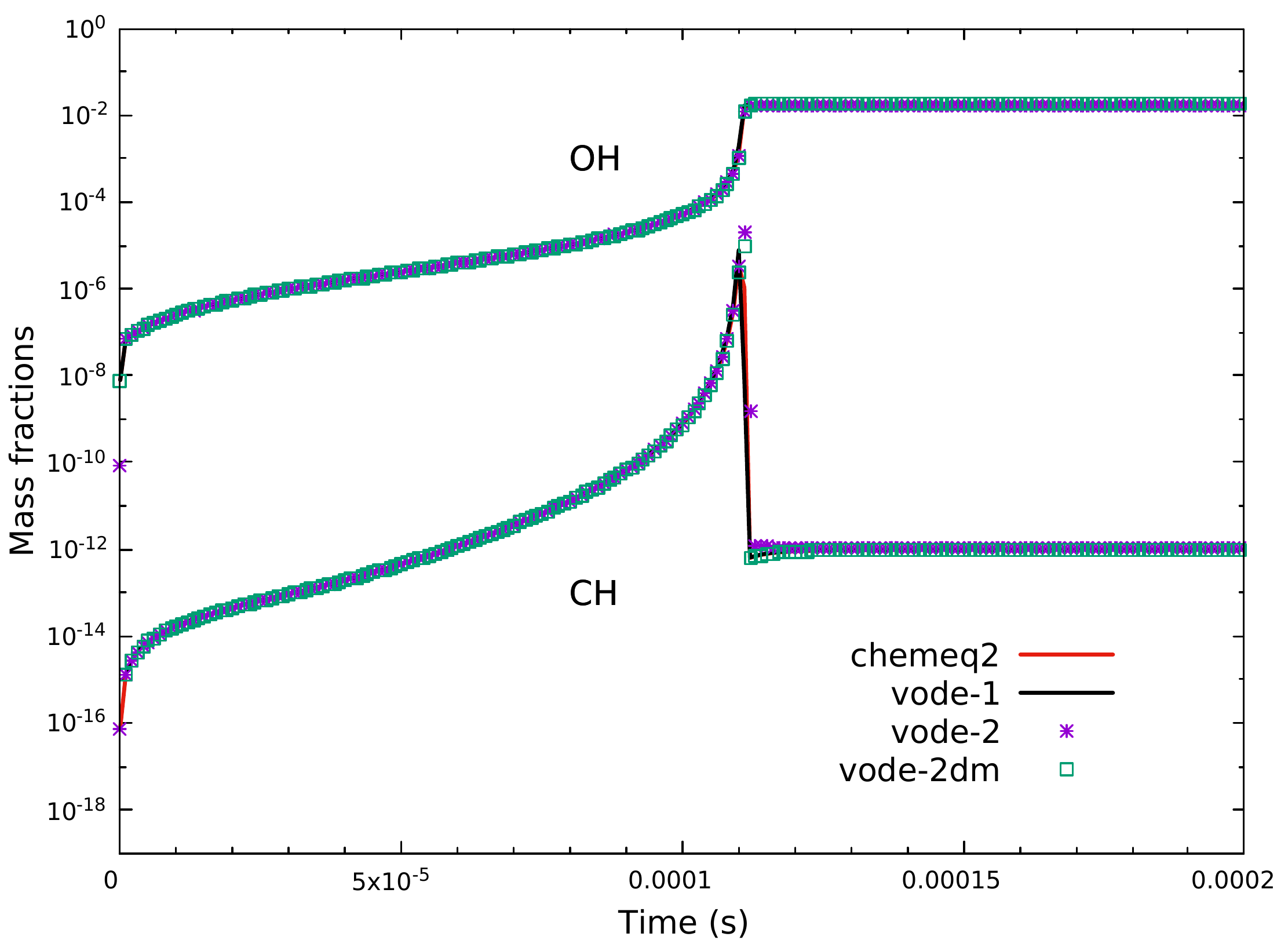} 
	\caption{ Calculated temperature and mass fraction histories for methane/air ignition delay problem in two initial conditions: left column (Case 1) and right column (Case 2). }    
	\label{case1_comparison}
\end{figure}

\begin{figure}
	\centering
	\includegraphics[scale=0.6]{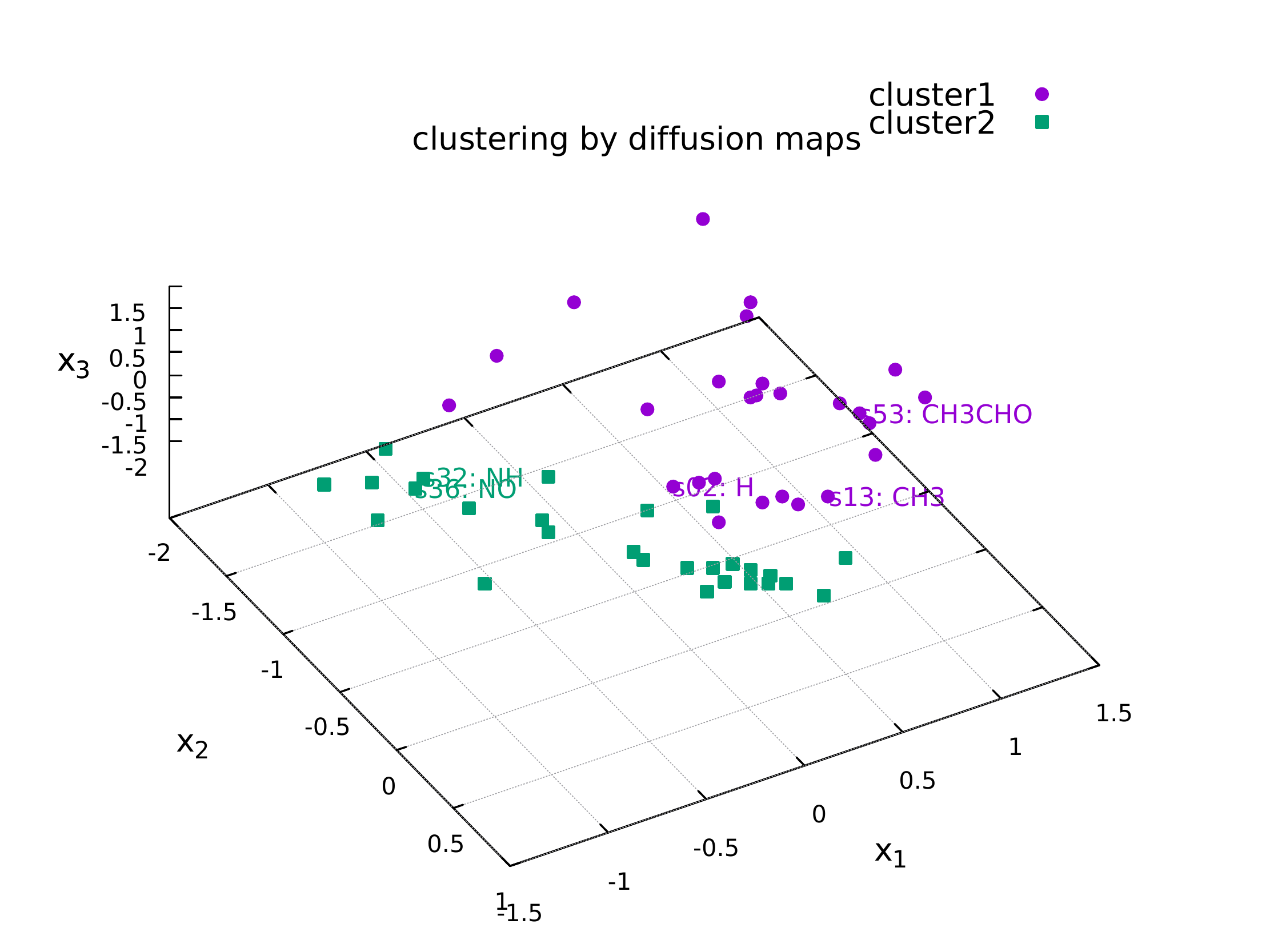}
	\includegraphics[scale=0.6]{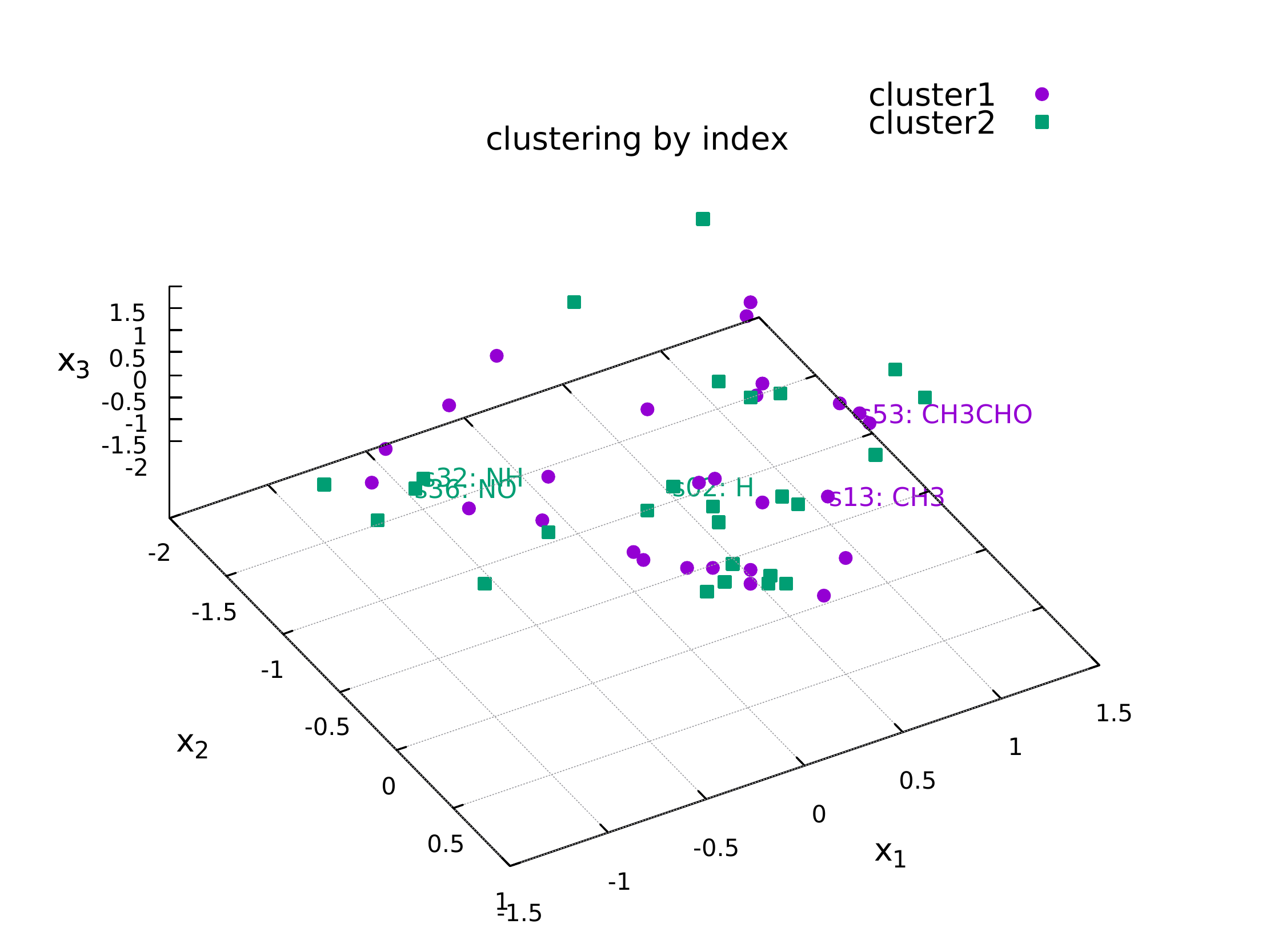} 
	\caption{ Embedding with first three diffusion coordinates of species for methane mechanism. }    
	\label{case1_clustering}
\end{figure}

\begin{figure}
	\centering
	\includegraphics[scale=0.32]{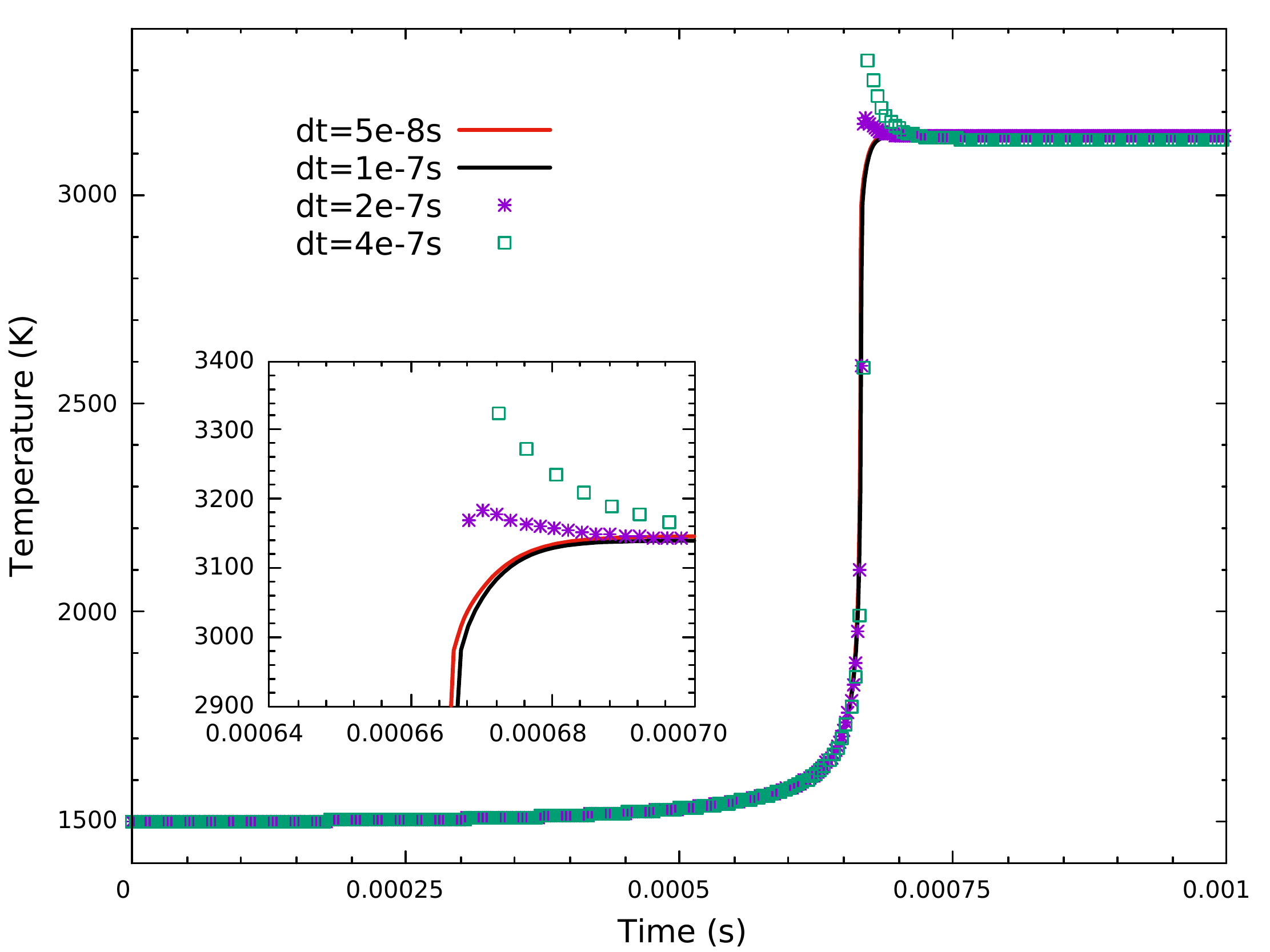}
	\includegraphics[scale=0.32]{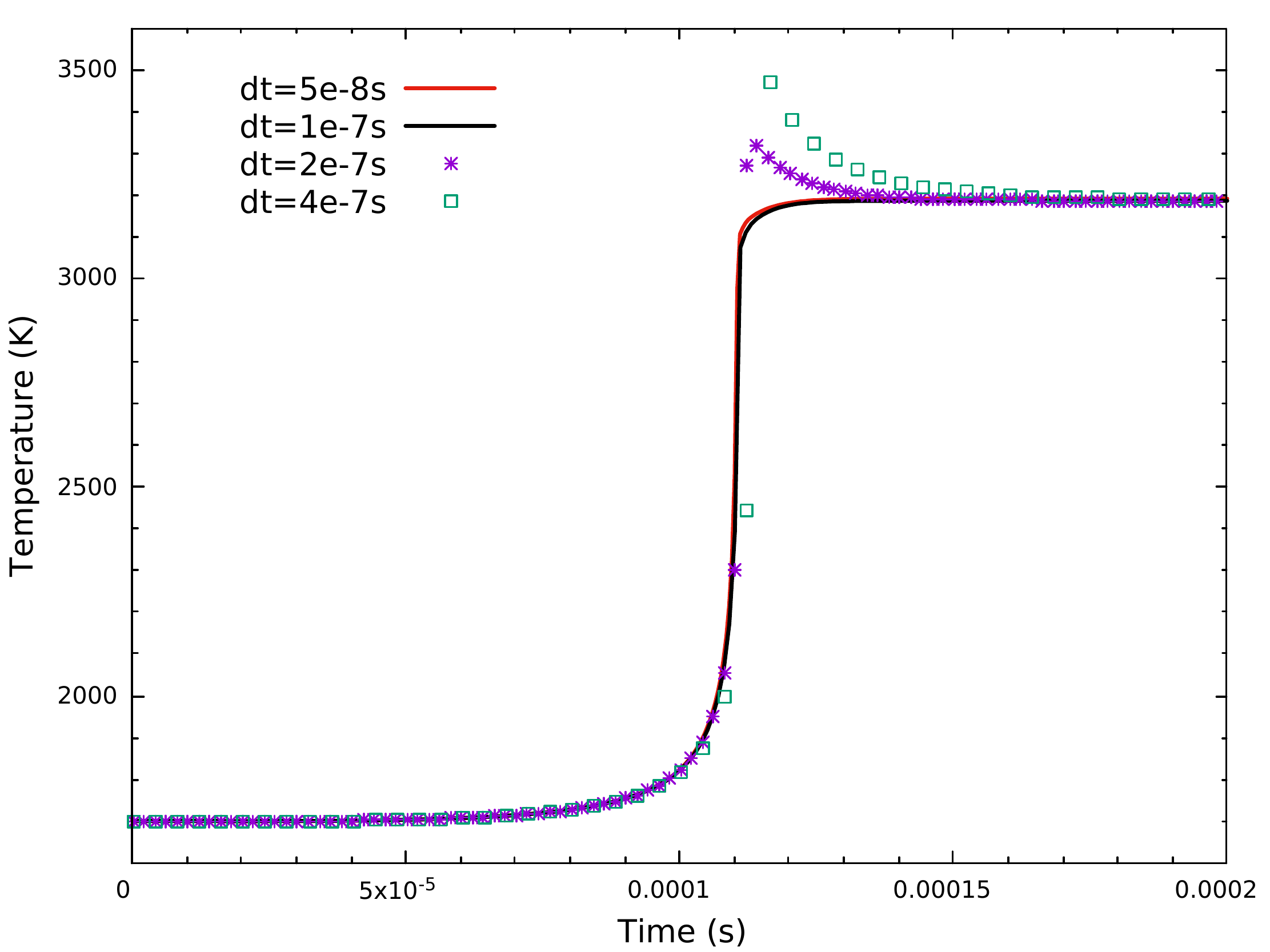} 
	\includegraphics[scale=0.32]{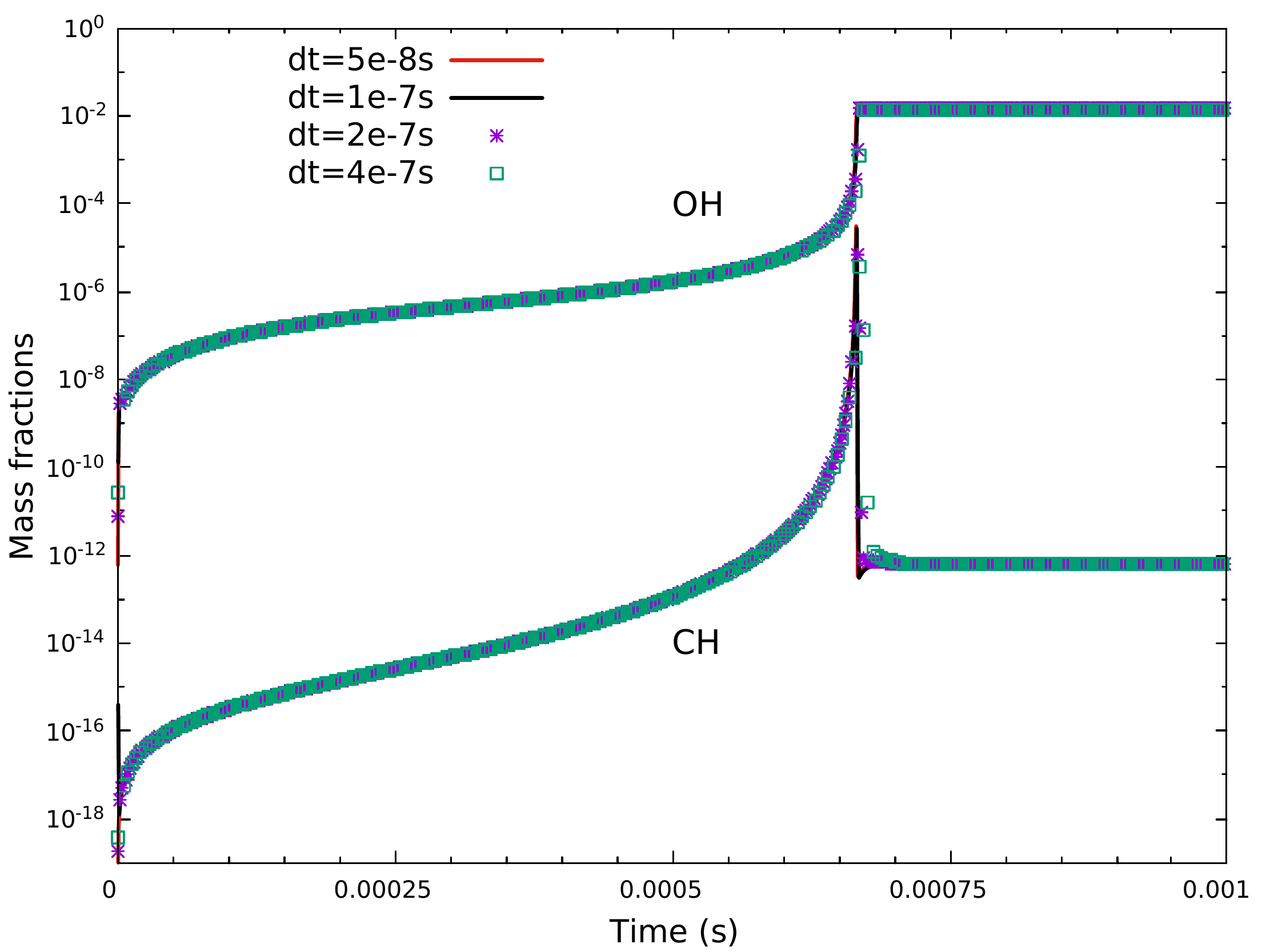}
	\includegraphics[scale=0.32]{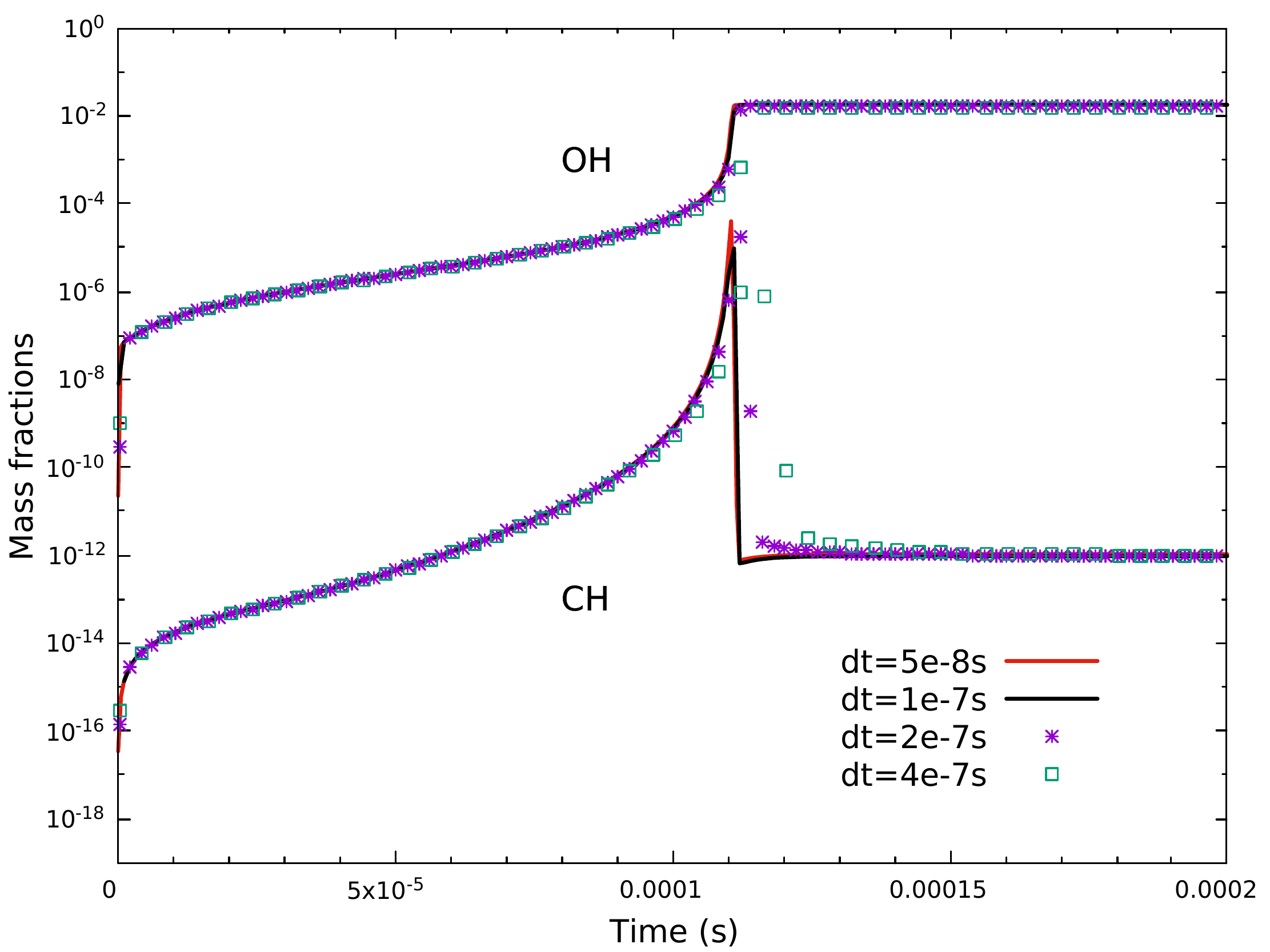} 
	\caption{ Calculated temperature and mass fraction histories for methane/air ignition delay problem by species clustering using varying timesteps in two initial conditions: left column (Case 1) and right column (Case 2). }    
	\label{case1_convergence}
\end{figure}

To validate operator splitting by species, the results obtained by CHEMEQ2 and VODE with/without species clustering are plotted in Fig. \ref{case1_comparison}, where VODE1 is without species clustering (that is, all the species are solved in only one set and a single step) while both VODE2 and VODE2-dm partition the species into two clusters for operator splitting by setting $N=2$. The difference of clustering lies in that VODE2 simply clusters the species in accordance with the species' index in the mechanism (e.g. species of odd or even indexing numbers are clustered in different subsets) while VODE2-dm utilizes diffusion maps for species clustering based on the weight matrix defined in Eq. \eqref{weight_matrix1}. In general, one clustering based on the indexing number of species can be readily obtained by
\begin{equation}\label{clustering_by_index}
\begin{aligned}
\text{Species } i \in
\begin{cases}
\text{cluster } 1: & \text{if } mod(i, N) = 1,\\
\text{cluster } 2: & \text{if } mod(i, N) = 2,\\   
  \cdots\\
\text{cluster } N-1: & \text{if } mod(i, N) = N-1,\\
\text{cluster } N: & \text{if } mod(i, N) = 0,\\
\end{cases}
&\end{aligned}
\end{equation}
where $i$ denotes the $i^{th}$ species in the mechanism and $N$ is the number of clusters by partition.
It can be seen that all the four solutions give the correct ignition delay times in two cases. In Case 1, VODE2 overestimates the temperature a little bit before it reaches an equilibrium state while VODE2-dm has almost the same temperature profile with both CHEMEQ2 and VODE1. The deficiency of VODE2 solution is amplified in Case 2, which also occurs at the end of the ignition process. Different predictions by VODE2 and VODE2-dm could be ascribed to the splitting error induced by partition: with diffusion maps, the quality of species clustering in VODE2-dm is better than that in VODE2. This can be illustrated by embedding the clustered species in a diffusion space, as shown in Fig \ref{case1_clustering}. 
As the clustered species are projected in the 3D diffusion space, we can clearly see that two clusters of species are separated from each other using diffusion maps, which indicates that each cluster is able to preserve the close interactions between coupling species. In particular, for the VODE2-dm clustering, 
the first species H and the last species CH3CHO are thrown into the same cluster as well as the 13th species CH3, due to the high activeness of H being involved in composition or decomposition reactions with hydrocarbon species such as
\begin{equation*}\label{clustering_by_index}
\begin{aligned}
\text{O}+\text{C}_2\text{H}_5 \Longleftrightarrow \text{H}+\text{CH}_3\text{CHO} ,\\
\text{H}+\text{CH}_3(+\text{M}) \Longleftrightarrow \text{CH}_4(+\text{M}). \\
\end{aligned}
\end{equation*}
Also, playing a critical role in the mechanism (as it participates in a large number of reactions), H is located at the center of diffusion space among all the species. On the other hand, species such as NO and NH are also reasonably clustered into the other subset because they mainly participate in nitrogen-related reactions, with looser interactions with hydrocarbon species. In contrast, H is clustered into the NH and NO group in the VODE2 clustering by index. Besides, the obtained two clusters mix with each other in the diffusion space and some pairs of two species with short distances are divided into different clusters, leading to the larger splitting error in VODE2 than that in VODE2-dm.
Furthermore, we examine convergence of the splitting method by varying the fixed timestep adopted in Fig. \ref{case1_convergence}. It is clear to see that as the timestep decreases the evolution of temperature and mass fractions approaches the corresponding profile at the shortest timestep: spikes in the temperature profiles with large timesteps gradually disappear and the jumps of mass fraction, $y_{CH}$, tend to sharpen by a sudden consumption during the ignition process.    
The base timestep of $\Delta t=1\times10^{-7}$ s is also verified to be sufficient for integrating the chemical kinetics correctly.                 
       
\subsection{n-Heptane/air auto-igniton case}

The second example considers the n-heptane/air combustion mechanism with a much larger scale. Two initial conditions \cite{yoo2011direct} are considered as in Table \ref{case34}. For Case 3, computation is carried out till $t=4 \times 10^{-4}$ s and the base timestep is also fixed at $\Delta t = 1 \times 10^{-7}$ s. Computation for Case 4 is till $t=1.1\times 10^{-4}$ s by 1100 steps. Without at-hand knowledge of the number of clusters which is most suitable and efficient for computing this large-scale mechanism, we choose to split the species by eight clusters using diffusion maps first.

\begin{figure}
	\centering
	\includegraphics[scale=0.32]{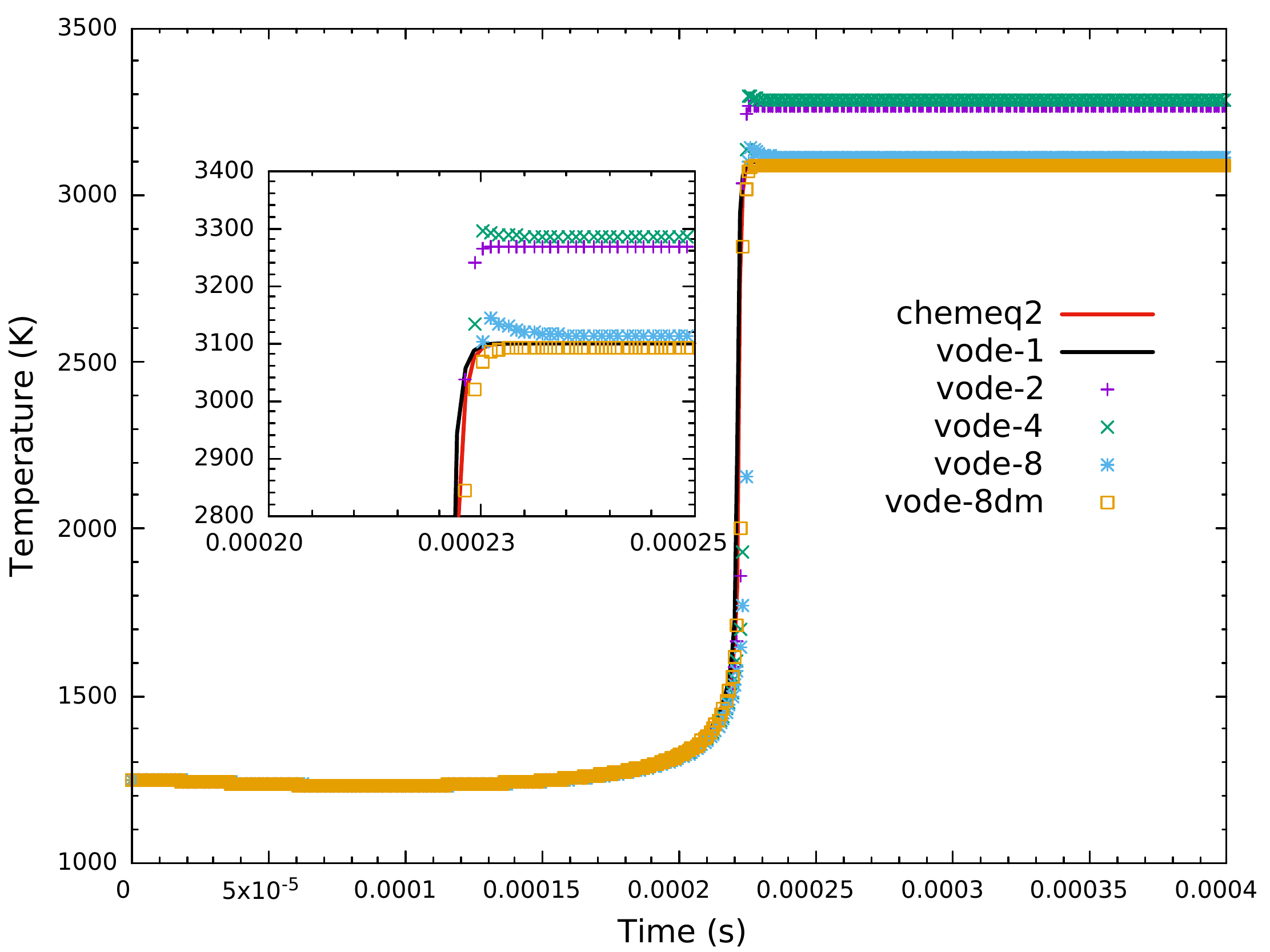}
	\includegraphics[scale=0.32]{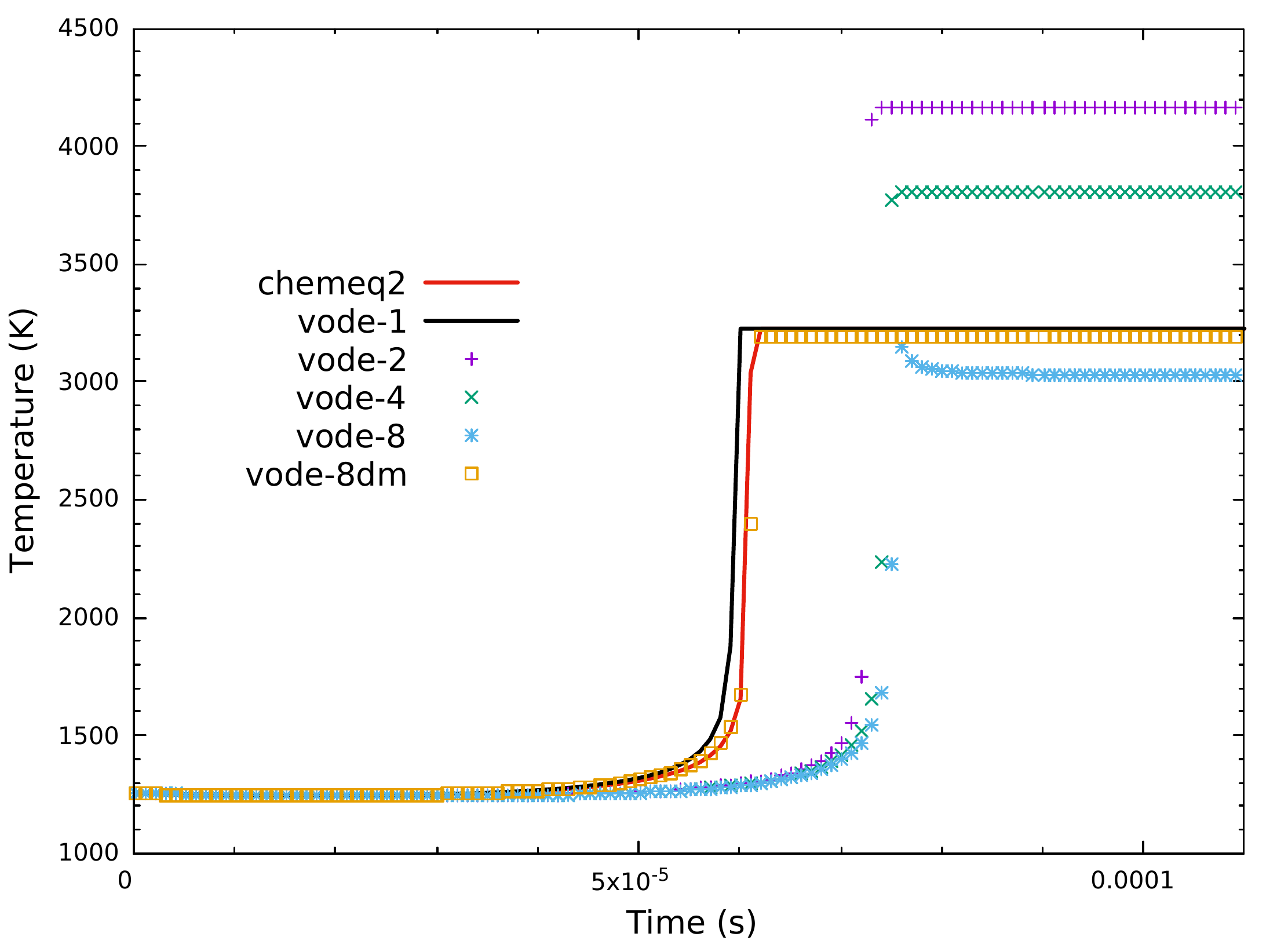} 
	\caption{ Calculated temperature histories for n-heptane/air ignition delay problem in two initial conditions: left column (Case 3) and right column (Case 4). }    
	\label{case2_comparison}
\end{figure}

In Fig. \ref{case2_comparison} the species clustered VODE result using diffusion maps is compared with those of simple clusterings using Eq. \eqref{clustering_by_index} by setting $N=2,4$ and 8, respectively, and also the results by CHEMEQ2 and non-split VODE. Calculated ignition delay times observed from the temperature histories of Case 3 and 4 by CHEMEQ2, VODE1 as well as VODE8-dm agree well with each other and also numerical results in Ref. \cite{yoo2011direct}. Using simple clustering algorithm instead of diffusion maps, VODE2, VODE4 and VODE8 obtain the correct ignition delay time for Case 3 while they all severely over-predict the delay of ignition for Case 4. Although the ignition delay time is not very sensitive to the species clustering in Case 3, the post-ignition equilibrium state appears to strongly depend on the quality of clustering as we can see that both VODE2 and VODE4 overestimate the equilibrium temperature incorrectly and VODE8 induces an incorrect spike before temperature reaches the equilibrium state, which is similar with the example of methane combustion. Besides, in Case 4, extremely high equilibrium temperatures nearly 4000 K and higher are induced by VODE2 and VODE4, and temperature spike also can be seen from the VODE8 solution.        

\begin{figure}
	\centering
	\includegraphics[scale=0.6]{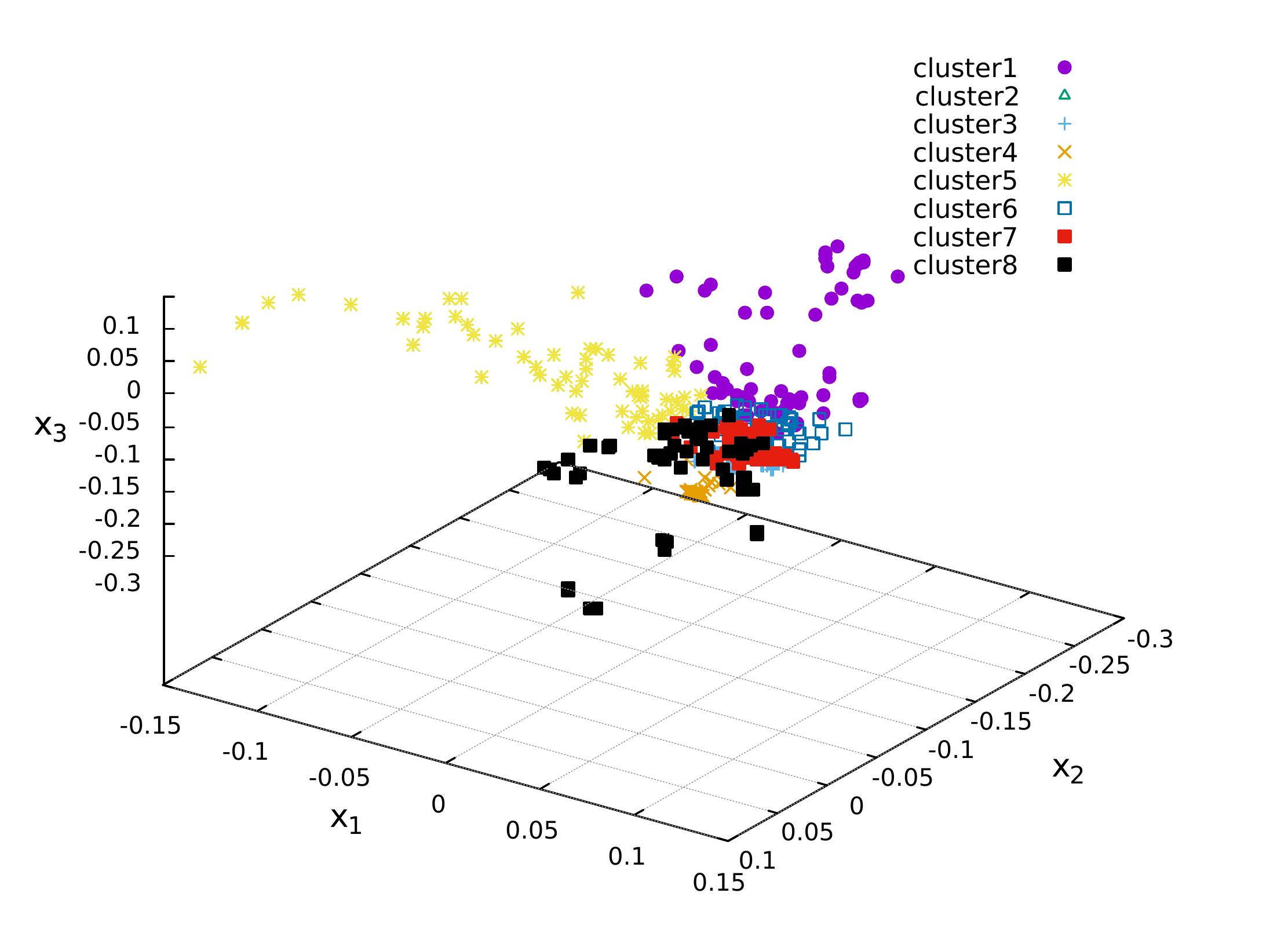} \\
	\includegraphics[scale=0.6]{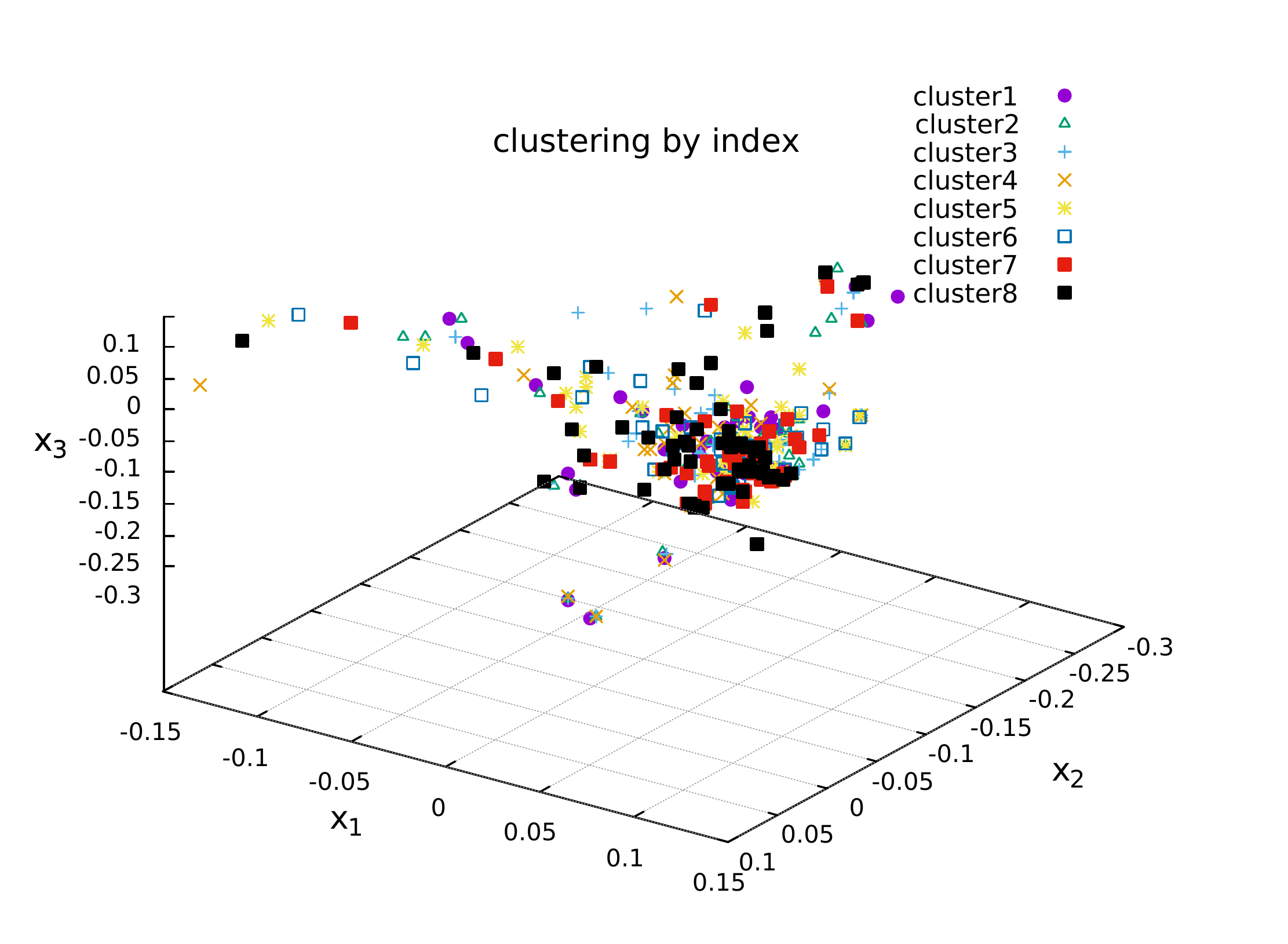}
	\caption{ Embedding with first three diffusion coordinates of species for n-heptane mechanism. }    
	\label{case2_clustering}
\end{figure}

\begin{figure}
	\centering
	\includegraphics[scale=0.32]{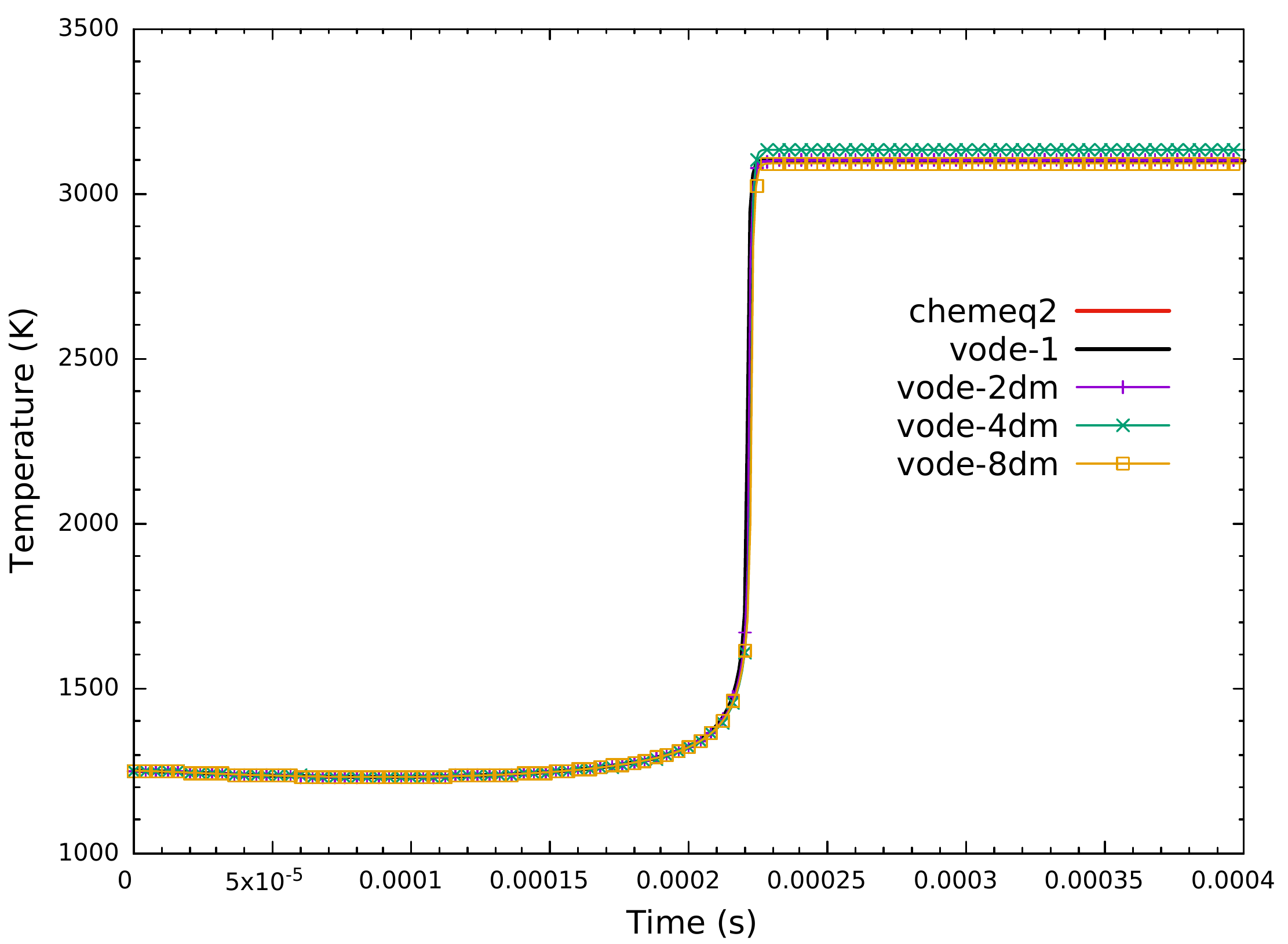}
	\includegraphics[scale=0.32]{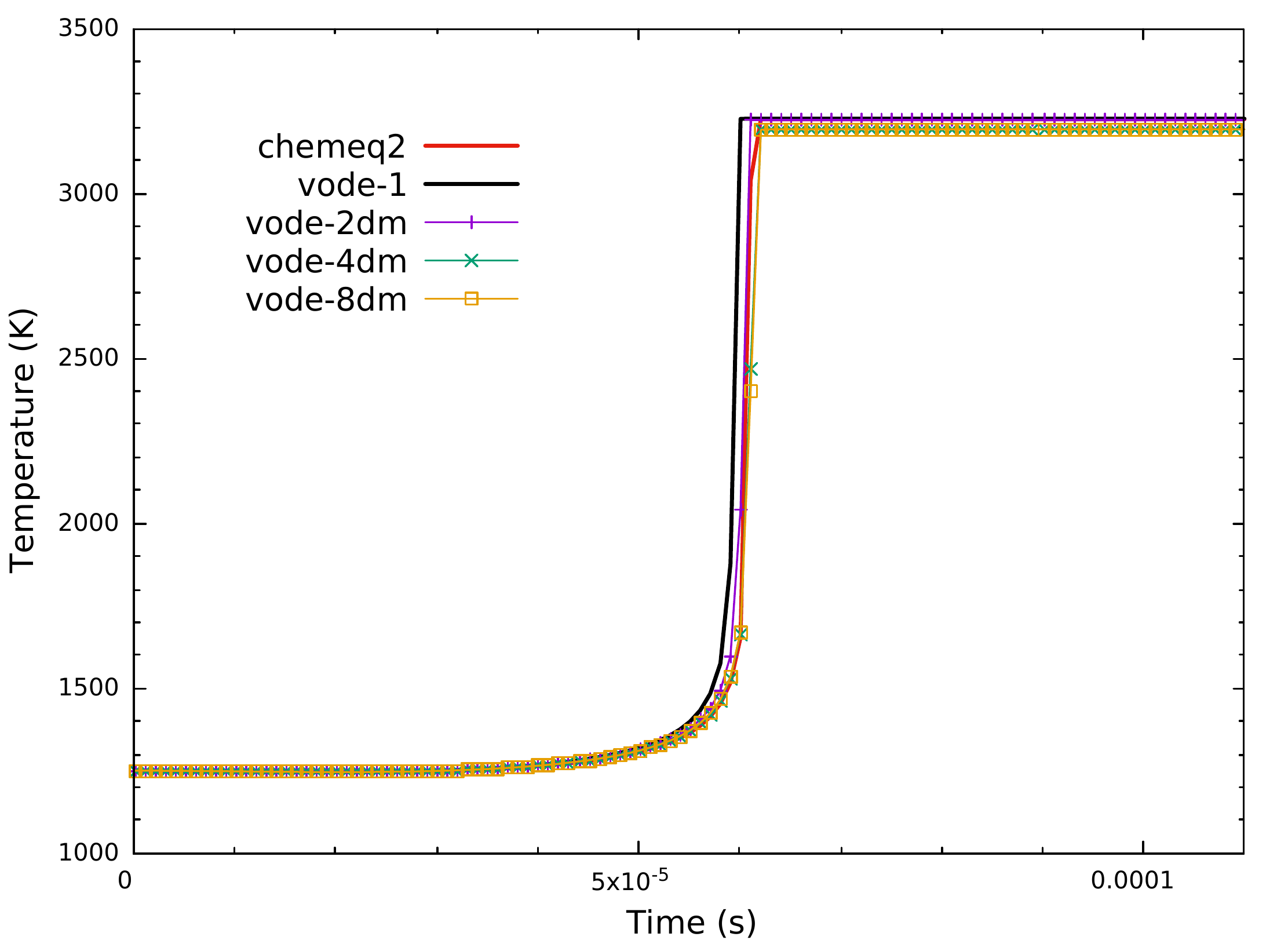} 
	\caption{ Calculated temperature histories for n-heptane/air ignition delay problem by species clustering setting $N=2,4,8$ in two initial conditions: left column (Case 3) and right column (Case 4). }    
	\label{case2_comparison_dm}
\end{figure}

In Fig. \ref{case2_clustering}, we present the species embedding with the first three coordinates, leading to eight clusters of species being scattered but well-organized in the diffusion space. In comparison, the simple clustering by index presents quite a disorder of species in the diffusion space. Quality of such a simple clustering is therefore expected to be poor, as shown in Fig. \ref{case2_comparison}.  
Besides, since the weight matrix is kept unchanged for the same mechanism, the diffusion space containing all the species is also the same and independent of the number of clusters one wants to partition. It is readily to further combine the close subsets (every two or four) into a larger cluster so that clustering by $N=4$ and $N=2$ can be straightforwardly obtained. Next, we compare the results denoted by VODE2-dm and VODE4-dm in Fig. \ref{case2_comparison_dm}. It can be seen that for both cases, the diffusion maps based results all capture the relatively correct ignition delay time and the equilibrium temperature. In particular, the VODE2-dm result performs better accuracy than the CHEMEQ2 result, being closer to the non-split VODE1 result. As the number of partition/splitting decreases, the split VODE results consistently approach the non-split solution, with reduced splitting errors.

\begin{figure}
	\centering
	\includegraphics[scale=0.45]{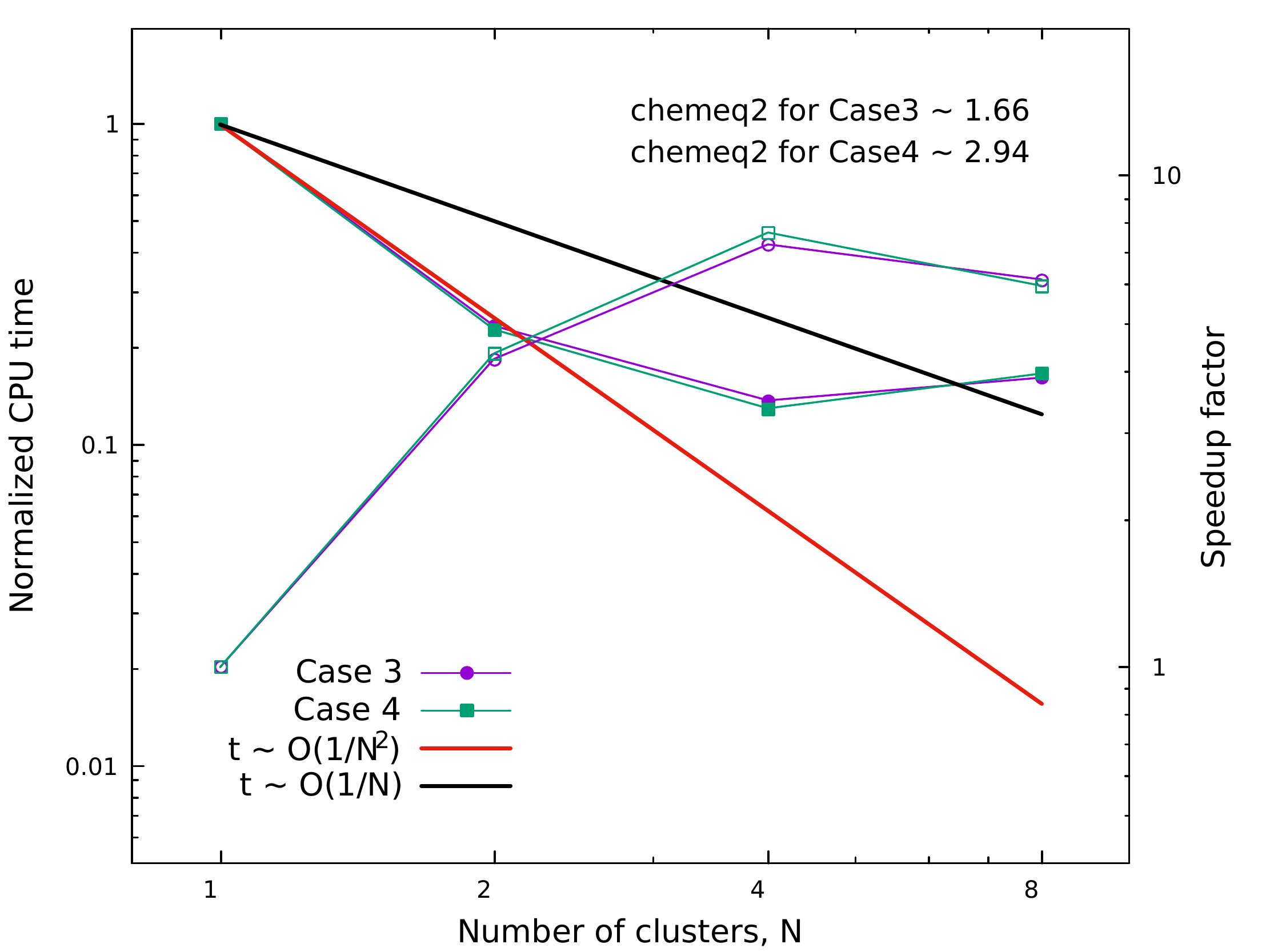}
	\caption{ Normalized CPU time and speedup factor by species clustered VODE with $N=1,2,4,8$; CPU time is normalized by $t/t_{vode1}$ and speedup factors use hollow symbols. 
	 }    
	\label{case2_cputime}
\end{figure}

In Fig. \ref{case2_cputime}, we investigate the computational efficiency of different solvers. All the results are normalized based on the CPU time of VODE1. It is to be noted that in these two cases, the non-split VODE solver runs faster than CHEMEQ2, indicating that implicit solvers are not necessarily slower than explicit solvers for large-scale problems. Focused on the split VODE solver using diffusion maps, we can see the declined CPU times as the number of clusters increases till $N=4$, falling within the zone bounded by two theoretical scalings according to Eq. \eqref{time_cost1}. When the number of clusters increases to $N=8$, the CPU time consumed meets a turning point and the computational efficiency is no longer monotonically decreasing. This is probably because, for a series of subsystems with a much smaller scale or dimension of matrix when $N \geq 8$, the time consumption of matrix operations such as the calculation of Jacobian matrix and LU factorization no longer dominate compared to that of calculating reaction rates and updating source terms. That is also to say, $N=4$ may be an optimal clustering number from the aspect of efficiency for the n-heptane ignition problem.                    

\begin{table}
	\caption{Initial conditions for n-heptane/air mixture.}
	\centering
	\label{case34}
	{
		\begin{tabular}{lccc}
			\hline
			&	n-C$_7$H$_{16}$:O$_2$:N$_2$ (mole)	& Temperature (K)	& Pressure (atm) \\
			\hline	
			Case 3	& \multirow{2}{*}{0.09091:1:3.76}	&	\multirow{2}{*}{1250}		&	10	\\
			Case 4 	& 							    	&								&	50	\\
			\hline
		\end{tabular}
	}
\end{table}

\subsection{n-Hexadecane/air auto-igniton case}

The third example considers the n-hexadecane/air combustion mechanism with the largest scale. Two initial conditions \cite{westbrook2009comprehensive} are considered as in Table \ref{case56}. For Case 5, computation is carried out till $t=1.1 \times 10^{-3}$ s while it is ceased for Case 4 at $t=2.2\times 10^{-4}$ s by 2200 equal timesteps. We also choose to split the species by eight clusters using diffusion maps first.

\begin{figure}
	\centering
	\includegraphics[scale=0.32]{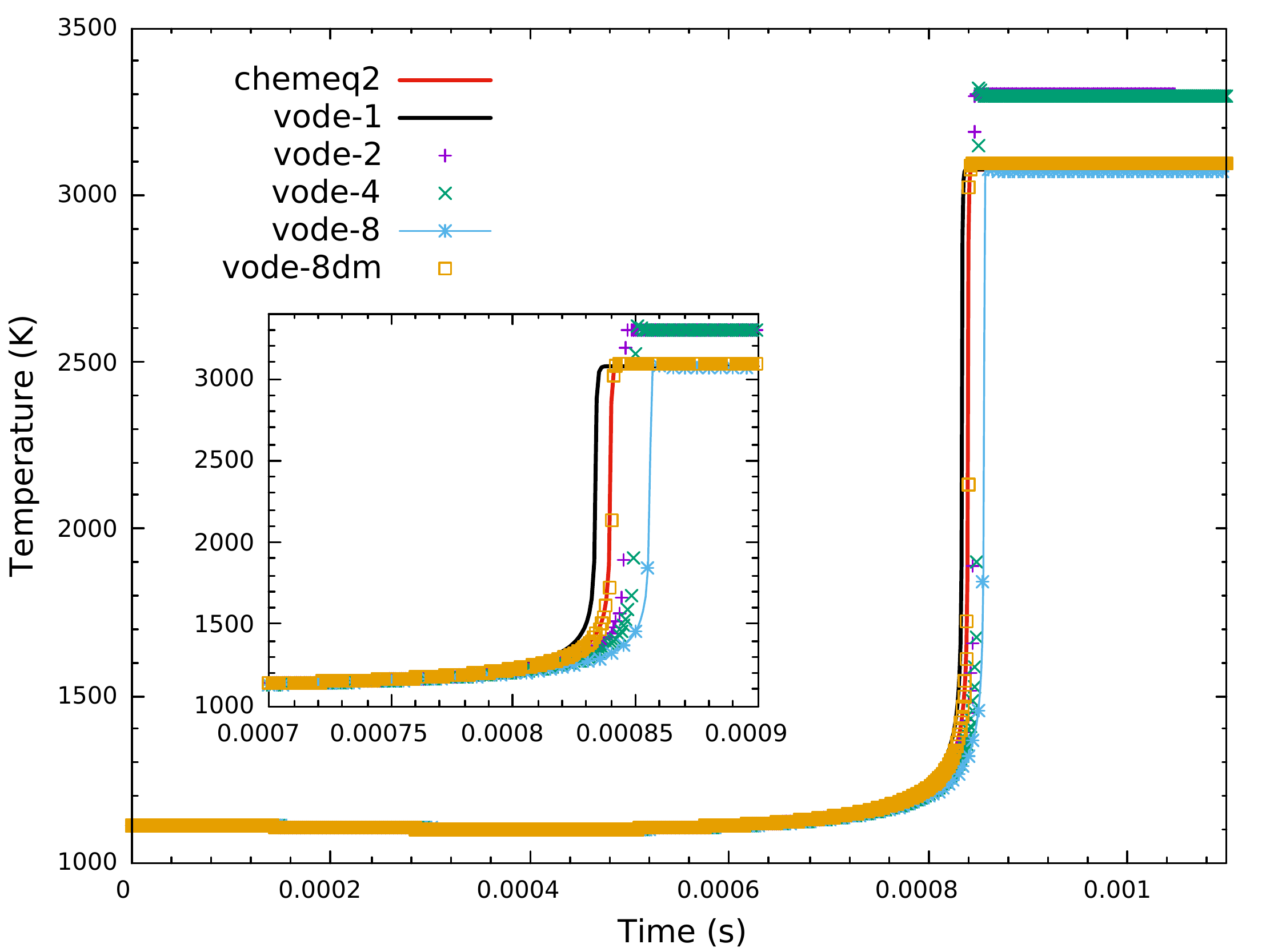}
	\includegraphics[scale=0.32]{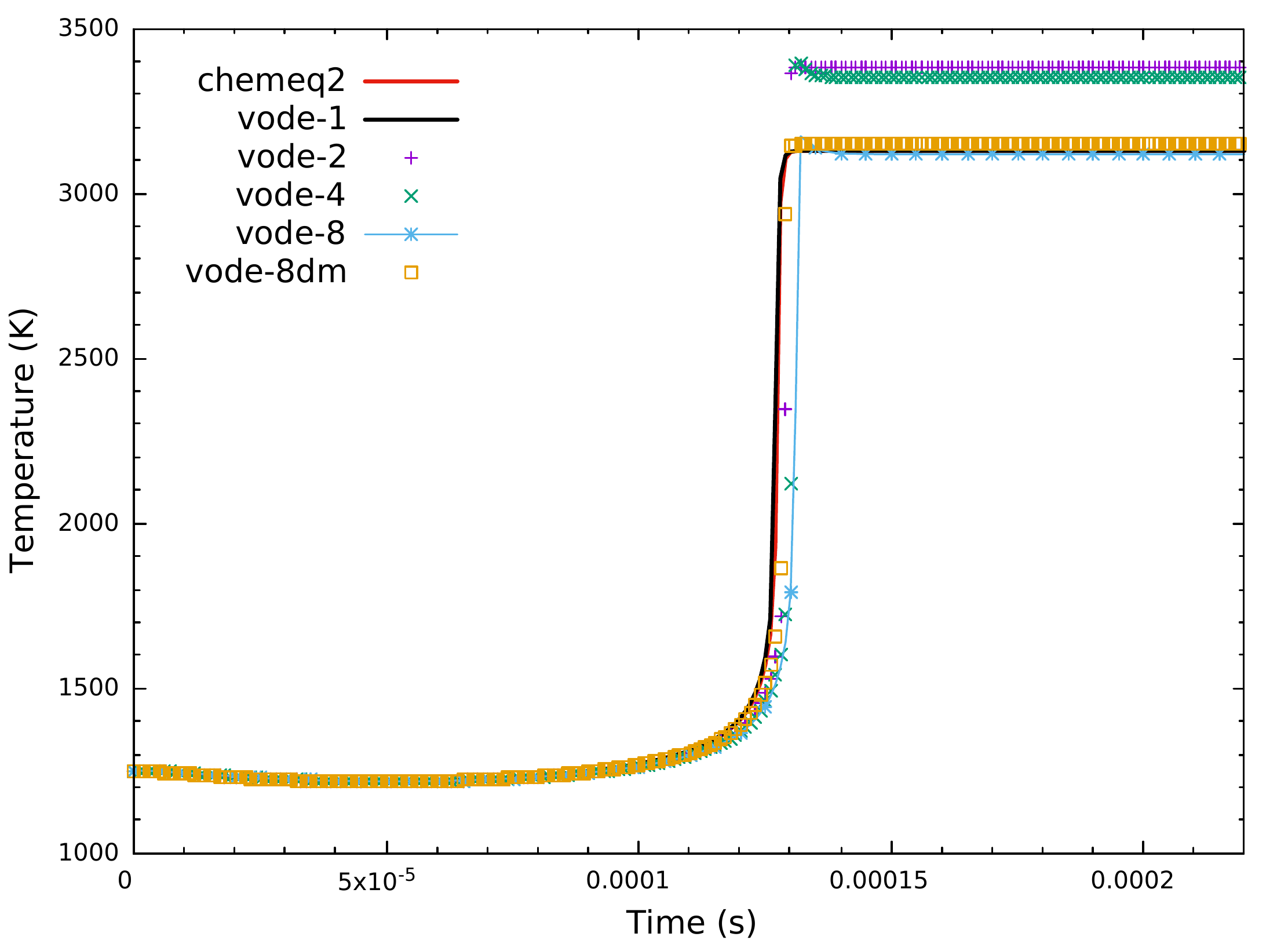} 
	\caption{ Calculated temperature histories for n-hexadecane/air ignition delay problem in two initial conditions: left column (Case 5) and right column (Case 6). }    
	\label{case3_comparison}
\end{figure}

In Fig. \ref{case3_comparison} the species clustered VODE result using diffusion maps is compared with those of simple clusterings using Eq. \eqref{clustering_by_index} by setting $N=2,4$ and 8, respectively, and also the results by CHEMEQ2 and non-split VODE. Calculated ignition delay times observed from the temperature histories of Case 5 and 6 by CHEMEQ2, VODE1 as well as VODE8-dm agree well with each other and also numerical results in Ref. \cite{westbrook2009comprehensive}. Using simple clustering algorithm instead of diffusion maps, VODE2, VODE4 and VODE8 obtain three increasing ignition delay times for Case 3 and 4. VODE8 computes the most delayed ignition time and both VODE2 and VODE4 overestimate the equilibrium temperature after ignition incorrectly. In contrast, the VODE8-dm result is comparable with the CHEMEQ2 result in both the ignition and post-ignition process.        

\begin{figure}
	\centering
	\includegraphics[scale=0.4]{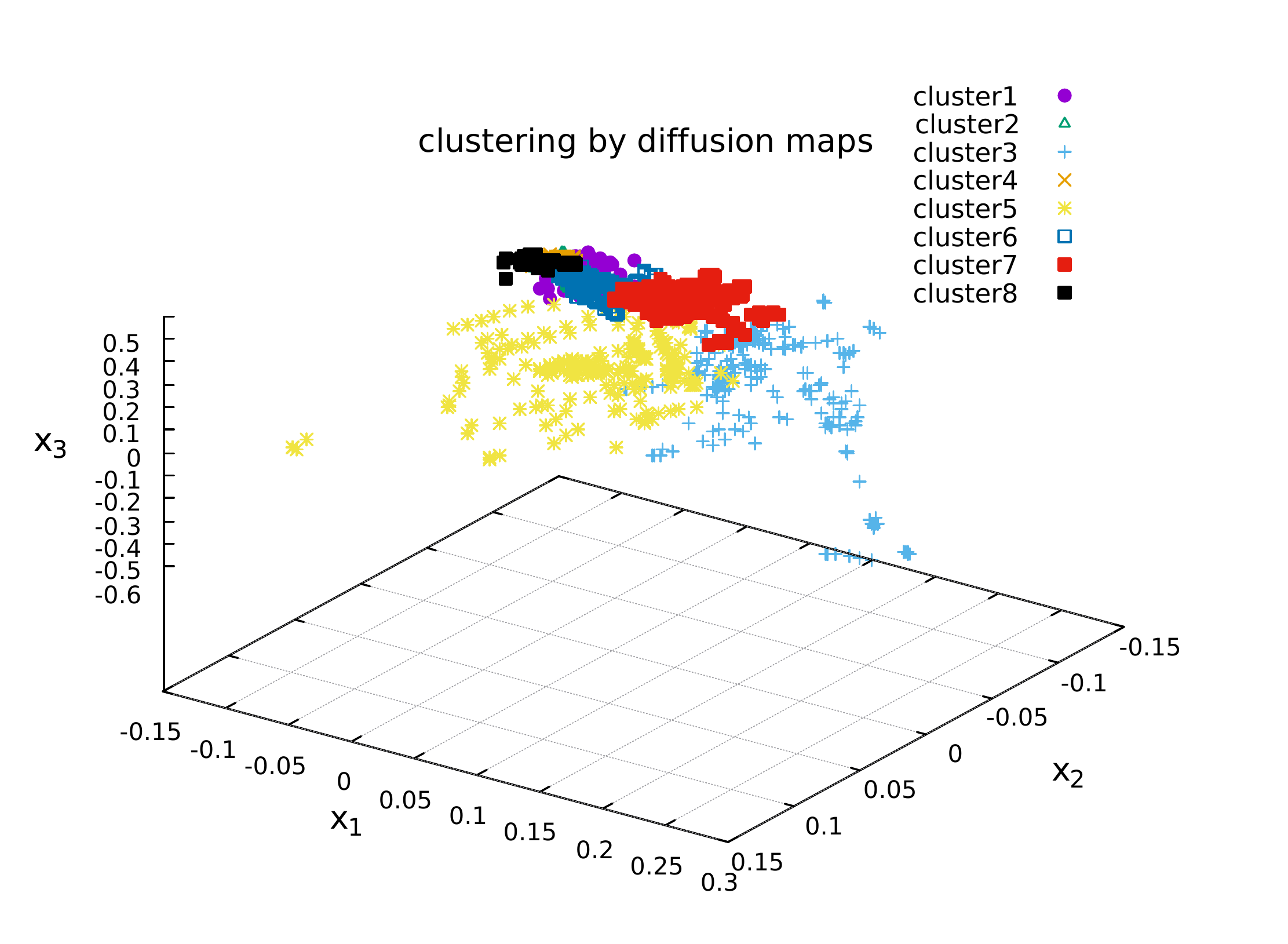} \\
	\includegraphics[scale=0.4]{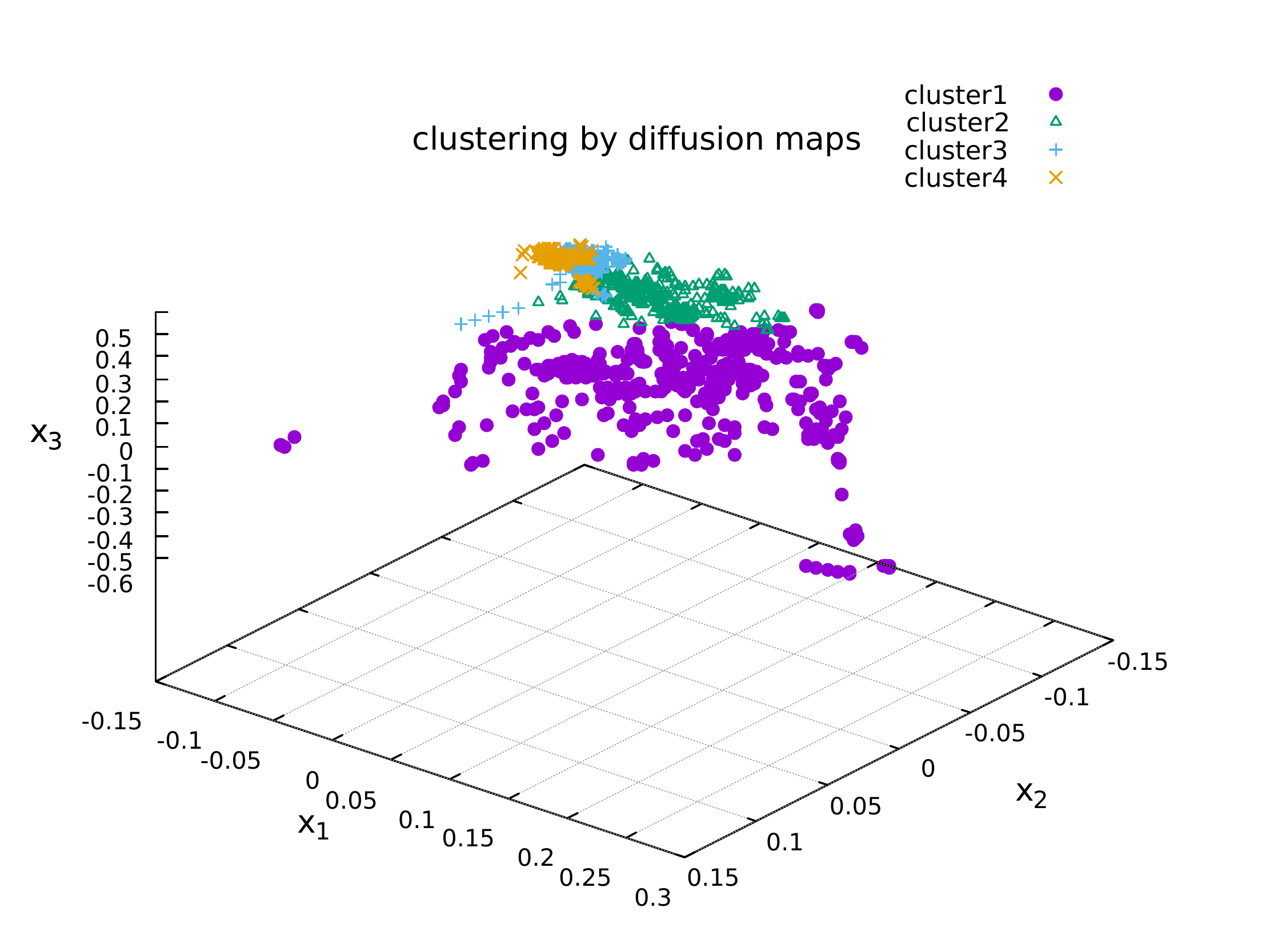} \\
	\includegraphics[scale=0.4]{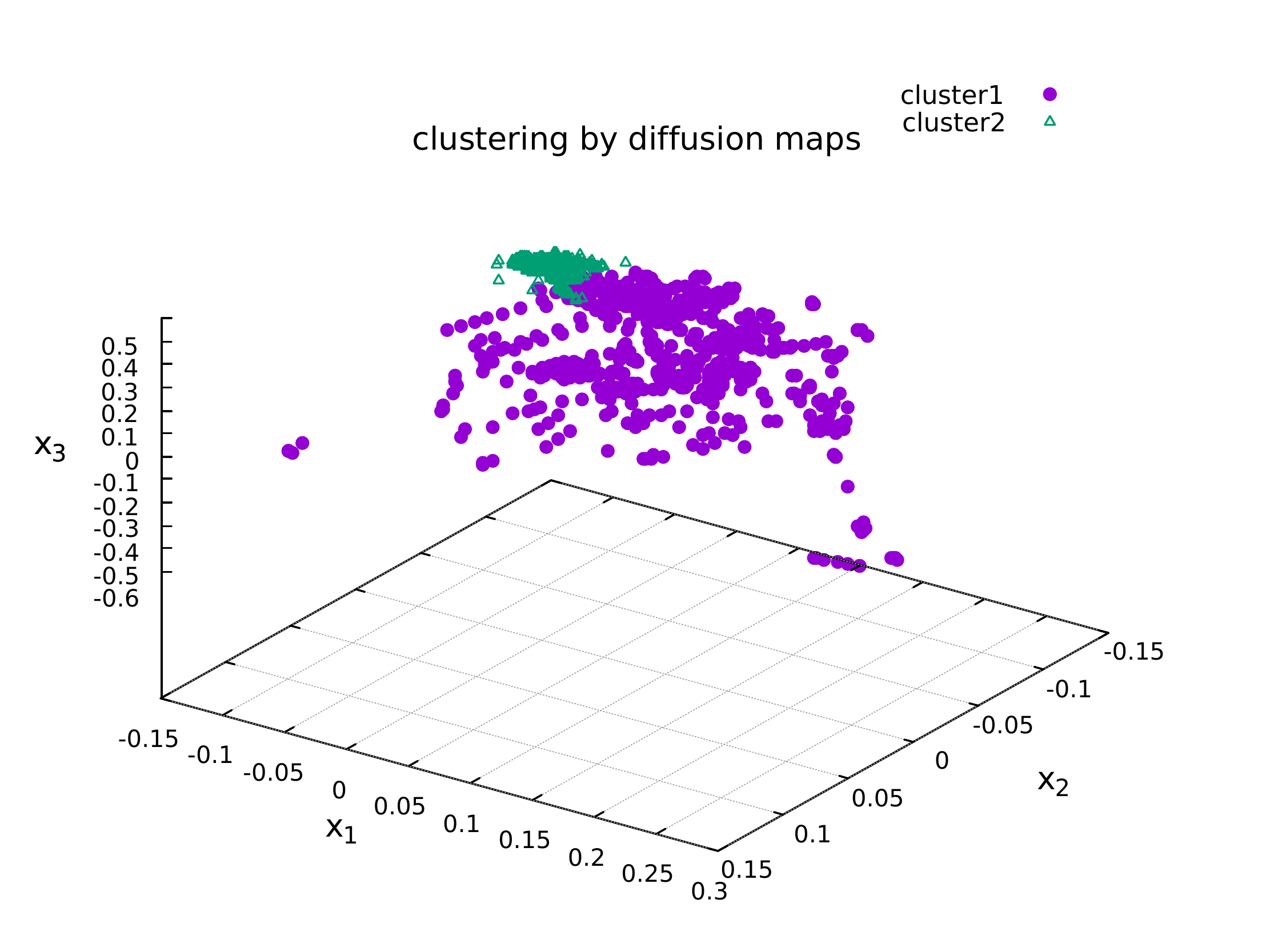} \\ 
	\caption{ Embedding with first three diffusion coordinates of species for n-hexadecane mechanism. }    
	\label{case3_clustering}
\end{figure}

\begin{figure}
	\centering
	\includegraphics[scale=0.32]{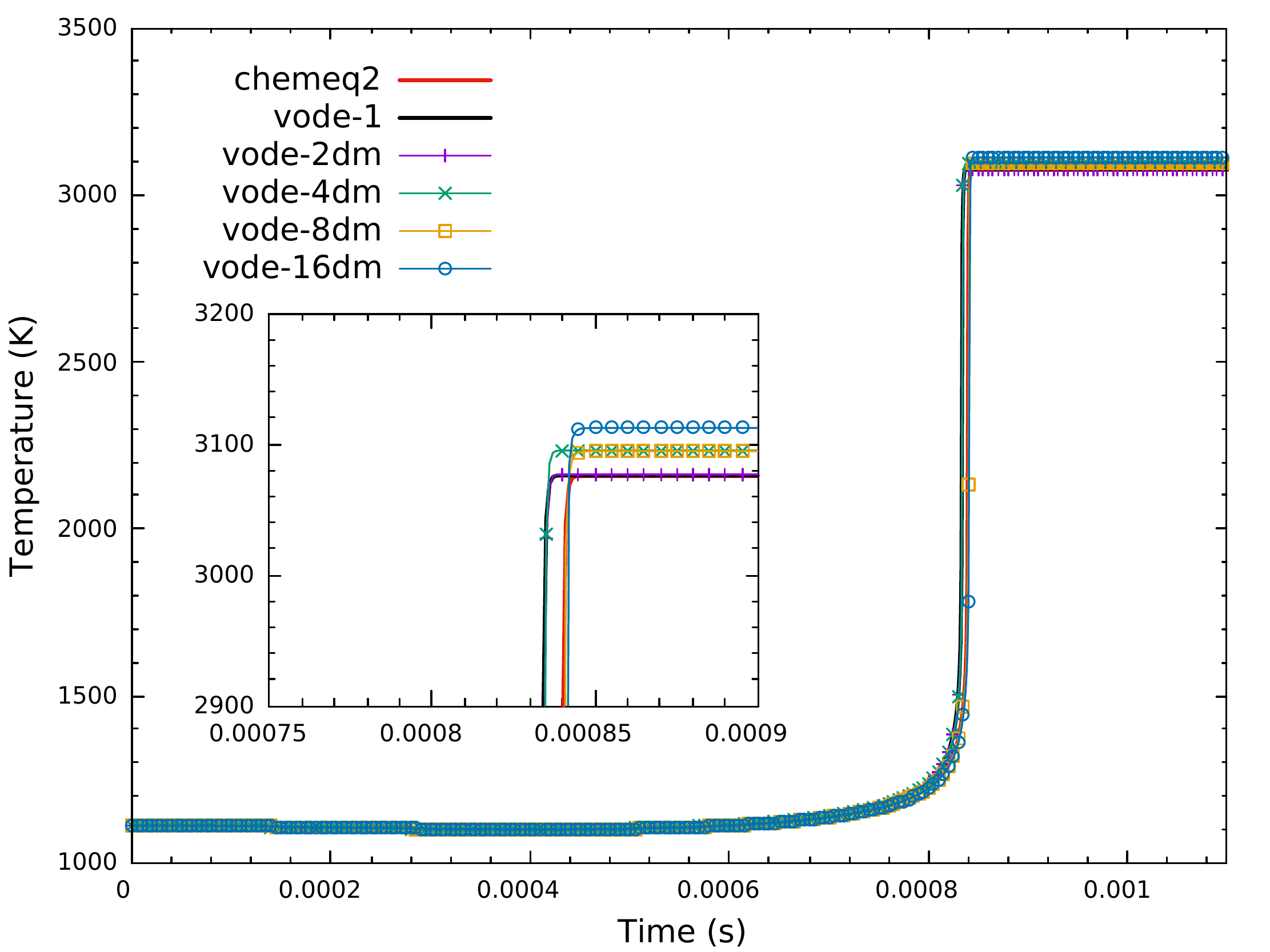}
	\includegraphics[scale=0.32]{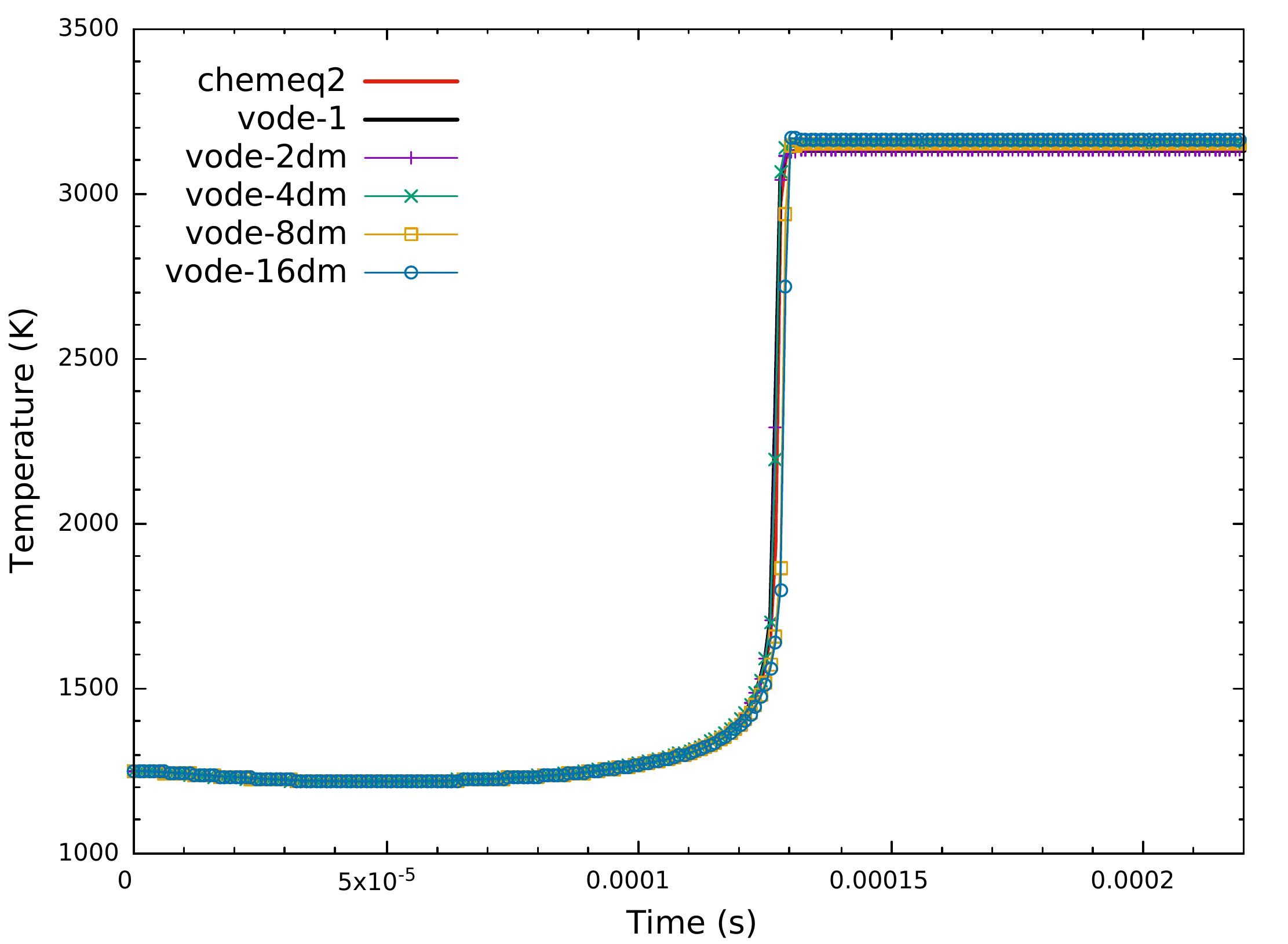} 
	\caption{ Calculated temperature histories for n-hexadecane/air ignition delay problem by species clustering setting $N=2,4,8,16$ in two initial conditions: left column (Case 5) and right column (Case 6). }    
	\label{case3_comparison_dm}
\end{figure}

In Fig. \ref{case3_clustering}, we present the species embedding with the first three coordinates, leading to eight/four/two clusters of species being scattered in the diffusion space. It is readily to see that the clustering with less number of clusters basically combines the close subsets of species into a larger cluster, as it is manually realized in the n-heptane example. By comparing the five results with difference number of clusters up to $N=16$ based on diffusion maps in Fig. \ref{case3_comparison_dm}, it is demonstrated that for both cases, the diffusion maps based results all capture the relatively correct ignition delay time and the equilibrium temperature. In particular, the VODE2-dm result performs the best, even better than the CHEMEQ2 result, being closest to the non-split VODE1 result. The VODE16-dm result slightly overestimates the equilibrium temperature and the ignition time predicted by VODE8-dm is later than that of VODE4-dm by $1 \times 10^{-5}$ s roughly. Though, as the number of partition/splitting decreases, the split VODE results consistently approach the non-split solution, with reduced splitting errors.

\begin{figure}
	\centering
	\includegraphics[scale=0.45]{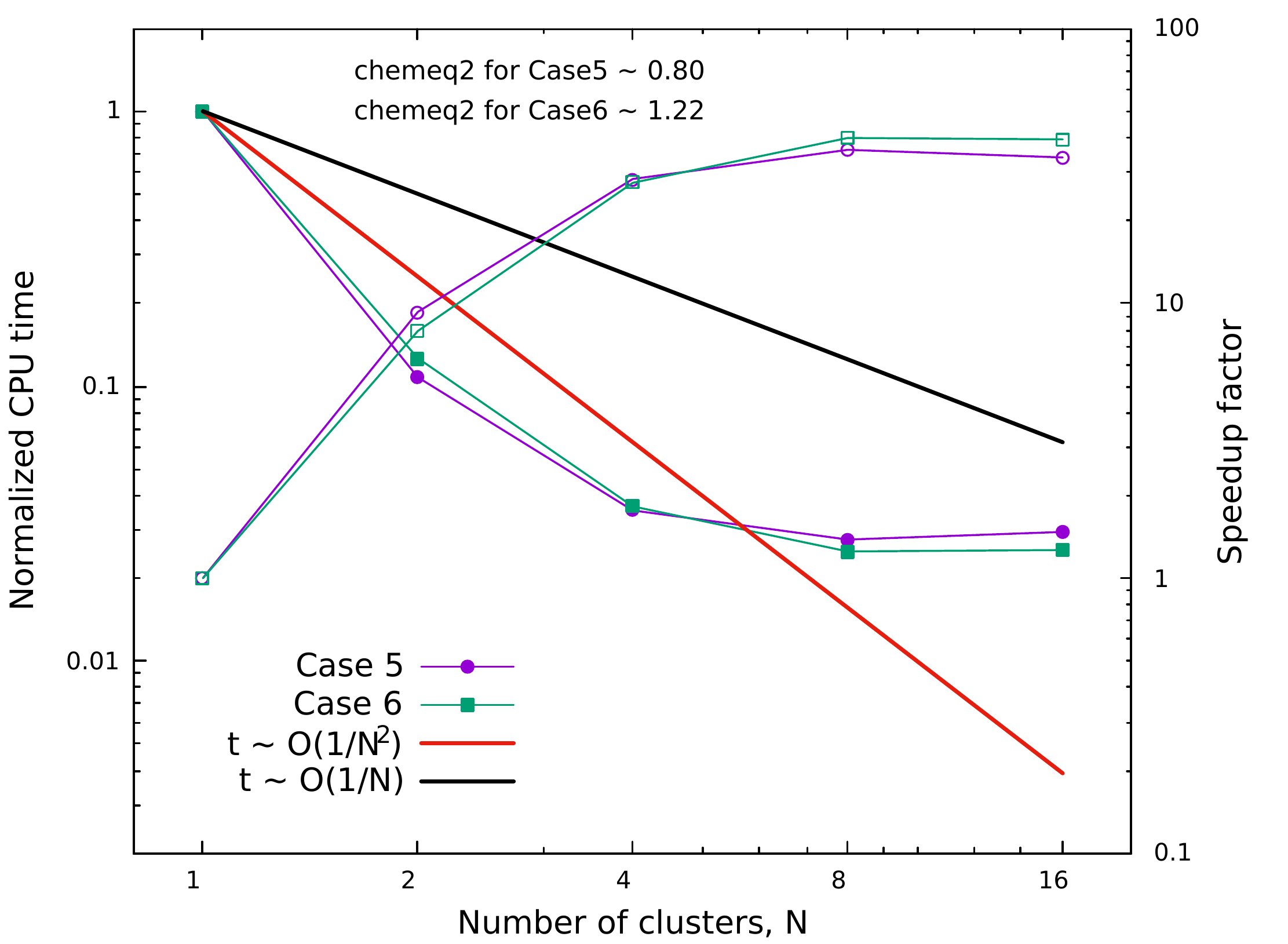}
	\caption{ Normalized CPU time and speedup factor by species clustered VODE with $N=1,2,4,8,16$; CPU time is normalized by $t/t_{vode1}$ and speedup factors use hollow symbols. 
	}    
	\label{case3_cputime}
\end{figure}

In Fig. \ref{case3_cputime}, we again investigate the computational efficiency of different solvers. It is to be noted that CHEMEQ2 is more efficient than VODE in the first case while in the second case the non-split VODE solver runs faster than CHEMEQ2, both solvers spending the CPU time with the same order of magnitude. Focused on the split VODE solver using diffusion maps, we can see the declined CPU times as the number of clusters increases till $N=8$ and performance in computation efficiency of the clustered VODE solvers when $N=2$ or $4$ even exceeds the theoretical expectation. It is probably because the consumption of CPU time such as in calculating reaction rates and updating source terms, etc, other than matrix operations is also dramatically reduced at a faster speed. When the number of clusters increases to $N=16$, the CPU time consumed no longer decreases sharply, indicating $N=8$ may be an optimal clustering number from the aspect of efficiency for the n-hexadecane ignition problem. It is to be noted that a total speedup factor of around 40 is realized by VODE8-dm for Case 5 and 6. It is even 50 times faster than CHEMEQ2 in the computation of Case 6, which is very promising for large-scale ODE systems without appealing to additional speedup treatments for matrix operations.

\begin{table}
	\caption{Initial conditions for n-hexadecane/air mixture.}
	\centering
	\label{case56}
	{
		\begin{tabular}{lccc}
			\hline
			&	n-C$_{16}$H$_{34}$:O$_2$:N$_2$ (mole) & Temperature (K)	& Pressure (bar) \\
			\hline	
			Case 5	& \multirow{2}{*}{0.04082:1:3.76} &	1111.11		&	\multirow{2}{*}{13.5}	\\
			Case 6 	& 							      &	1250		&							\\
			\hline
		\end{tabular}
	}
\end{table}

\section{Conclusions}
\label{Conclusions}

For large-scale chemical kinetics involving many species and reactions, computational efforts needed for time integration are usually far more than linearly depending on size of the kinetic mechanism, especially when implicit ODE solvers are used. To archive computational speedup, in this paper we utilize operator splitting to integrate the large system in separate yet consecutive subsystems of the same and smaller size. Each subsystem includes a cluster of species decoupled from other species in the whole mechanism and is solved separately and implicitly by VODE. In order to reduce the inevitable splitting error, diffusion maps are applied to analyze the species graph and cluster strongly coupled species into the same subsystem, by defining an appropriate weight matrix for chemical kinetics. Three hydrocarbon fuel/air ignition problems with an increasing scale of the mechanism, up to 2115 species and 8157 reactions, are taken into consideration under varying initial conditions. 
Computational efficiency and accuracy can be reasonably compromised by choosing a proper number of clusters using diffusion maps. For the n-heptane mechanism, partition by 4 clusters of species leads to about 8 times speedup compared to the non-split VODE solver and $10 \sim 20$ times speedup versus the explicit solver CHEMEQ2. For the n-hexadecane mechanism, partition by 8 clusters of species results in a speedup factor of XXX. Clustering by diffusion maps based on the present weight matrix outperforms the simple clustering according to species' index in the mechanism in terms of capturing the correct ignition delay time and post-ignition equilibrium state. It implies that an optimal clustering for a certain mechanism is profitable not only for computational acceleration but also for less sacrifice of accuracy.    
Therefore, either the new definition of weight matrix or other optimal partition techniques in addition to diffusion maps is worthy of further comprehensive investigation in our future work. Besides, the above splitting is conducted in accordance with the first-order Lie-Trotter scheme and extension to higher-order schemes which may allow for a greater number of clusters is straightforward.

\section*{Acknowledgements}
The financial support from the EU Marie Sk{\l}odowska-Curie Innovative Training Networks (ITN-ETN) (Project ID: 675528-IPPAD-H2020-MSCA-ITN-2015) for the first author is gratefully acknowledged. Xiangyu Y. Hu is also thankful for the partial support by National Natural Science Foundation of China（(NSFC) (Grant No:11628206).

\newpage
\section*{References}
\bibliographystyle{abbrv}
\bibliography{elsarticle-template-num}

%
%
%
%

\end{document}